\documentclass[fleqn,usenatbib,manuscript]{mnras}

\usepackage{newtxtext,newtxmath}
\usepackage[T1]{fontenc}
\usepackage{ae,aecompl}
\usepackage{graphicx}	% Including figure files
\usepackage{amsmath}	% Advanced maths commands
\usepackage{amssymb}	% Extra maths symbols
\usepackage{soul,xcolor}

\newcommand{\noopsort}[1]{}

\title[Particle transport in shock simulations]{Particle transport in hybrid PIC shock simulations: A comparison of diagnostics}

\author[Trotta et al.]{
D. Trotta $^{1}$\thanks{E-mail: d.trotta@qmul.ac.uk}
, D. Burgess $^{1}$
, G. Prete $^{2}$
, S. Perri $^{2}$ 
and G. Zimbardo $^{2}$
\\
$^{1}$Queen Mary University of London, School of Physics and Astronomy, London E1 4NS, UK \\
$^{2}$Dipartimento di Fisica, University of Calabria, Italy \\
}

\date{ Accepted 2019 September 26. Received 2019 September 25; in original form 2019 March 4}

\pubyear{2019}
\begin{document}
\label{firstpage}
\pagerange{\pageref{firstpage}--\pageref{lastpage}}
 \newcommand{\quomark}[1]{``#1''}

\maketitle

\begin{abstract}
Recent \emph{in-situ} and remote observations suggest that the transport regime associated with shock accelerated particles {may} be anomalous {i.e., the Mean Square Displacement (MSD) of such particles scales non-linearly with time}. We use self-consistent, hybrid PIC plasma simulations to simulate a quasi-parallel shock {with parameters compatible with heliospheric shocks,} and gain insights about the particle transport in such a system. For suprathermal particles interacting with the shock we compute the MSD separately in the upstream and downstream regions. Tracking suprathermal particles for sufficiently long times up and/or downstream of the shock poses problems in particle plasma simulations, such as statistically poor particle ensembles and trajectory fragments of variable length in time. Therefore, we introduce the use of time-averaged mean square displacement (TAMSD), which is based on single particle trajectories, as an additional technique to address the transport regime for the upstream and downstream regions. MSD and TAMSD are in agreement for the upstream energetic particle population, and both give a strong indication of superdiffusive transport, consistent with interplanetary shock observations. MSD and TAMSD are also in reasonable agreement downstream, where indications of anomalous transport are also found. TAMSD shows evidence of heterogeneity in the diffusion properties of the downstream particle population, ranging from subdiffusive behaviour of particles trapped in the strong magnetic field fluctuations generated at the shock, to superdiffusive behaviour of particles transmitted and moving away from the shock.
\end{abstract}


\section{Introduction}

High energy particle acceleration is a common feature of many astrophysical systems, and characterizing the energetic particle transport is a key challenge for understanding the underlying physical acceleration processes. In most cases, when particle trajectories are available, the method of mean square displacement (MSD) is the most widely used diagnostic to study particle transport. MSD uses ensemble averages, and hence can be unreliable with poor particle statistics. On the other hand, the method of time-averaged mean square displacement (TAMSD) relies on time  rather than ensemble averages, and is a popular diagnostic in research communities concerned with particle transport in biological systems. When dealing with particle tracking in self-consistent plasma simulations of acceleration it can be challenging to obtain sufficiently good particle statistics over long times, due to the non-periodicity and limited extent of such simulations and consequently the varying lifetimes of different particles in the simulation. For this reason, we investigate the use of TAMSD diagnostics to study the transport of accelerated particles in self-consistent hybrid PIC shock simulations, and make comparisons with the results from the MSD method.

It is widely accepted that shocks are efficient particle accelerators in astrophysical systems \citep[e.g.,][]{Blandford1987, Fisk2015}. Generally speaking, shocks convert directed flow energy (upstream) to thermal energy (downstream). In collisionless shocks, a small fraction of the upstream energy can be channeled towards particle acceleration. The mechanisms by which particles are accelerated out of the thermal pool and to high energies are still debated. One of the most important parameters that controls shock structure and behaviour in a collisionless plasma is the angle between the normal to the shock surface and the upstream magnetic field, $\theta_{Bn}$. Shocks with $\theta_{Bn}$ close to zero are called quasi-parallel, and they are believed to be more efficient at particle acceleration than quasi-perpendicular ones \citep[see, for example,][]{burgess_book}. Other important parameters are the shock Alfv\'enic and sonic Mach numbers, defined as $M_A = v_\mathrm{sh}/{v_A}$ and $M_S = v_\mathrm{sh}/{c_s}$, respectively,  and the plasma $\beta = v_\mathrm{th}^2/{v_A^2}$. Here, $v_A$ is the Alfv\'en speed in the region upstream from the shock, $v_\mathrm{sh}$ is the shock speed in the upstream flow frame, $c_s$ is the sound speed of the upstream flow and $v_\mathrm{th}$ is the thermal speed of the upstream flow.

Particle acceleration at shocks is invoked to explain a wide variety of direct and indirect observations. Direct evidence of shock accelerated particles has been found by means of spacecraft measurements since the 1960s \citep[e.g,][]{Bryant1962,Scholer1982,nes1984} for interplanetary shocks. At the solar wind termination shock, the suprathermal particle population was found enhanced by \emph{Voyager} spacecraft observations \citep[][]{Decker2005,Decker2008}. Indirect evidence of accelerated particles is commonly found in the X-ray signature of supernova remnants \citep[see][for a review]{reynolds2008}. Recently, particle acceleration at shocks has also been found to be relevant in the galaxy intracluster medium \citep[e.g.][]{vanweeren2010,vanweeren2017,Kang2017}.

A widely used theory for particle acceleration at quasi-parallel shocks is Diffusive Shock Acceleration theory (DSA) \citep[][]{Krymskii1977,Axford1978,Bell1978,Blandford1978,Drury1983,Jokipii1987}. The idea underlying DSA theory is that particles with Larmor radii larger than the shock thickness interact with the shock by diffusing upstream and downstream, and thus are continuously returned to the shock, gaining energy at each crossing. This process is often called first order Fermi acceleration. The main prediction of DSA is a power-law distribution in energy for the accelerated particles, with a slope that depends only on the shock compression ratio, at least in the simple time stationary limit. DSA theory also predicts that, in the case of a spatially constant diffusion coefficient, the density profiles of energetic particles should decrease exponentially upstream of the shock, and to be constant downstream. These predictions have been found to be not inconsistent with some observations of interplanetary shocks \citep[e.g.][]{Giacalone2012}, even though many \emph{in-situ} spacecraft crossings exhibit more complex energetic particle profiles \citep[see][]{nes1984}.

Recently, some energetic particle observations {have suggested that the transport regime associated with such particles can be anomalous}. In fact, energetic particle density profiles at interplanetary shocks are often found to show a power-law decay rather than an exponential one, as one would expect {from normal diffusion} \citep[][]{Perri2007,Perri2008a, Perri2009a, Perri2009b, Sugiyama2011,Perri2015}. In addition,  it has been proposed, among other possible mechanisms,  that the discrepancy between Mach numbers obtained from radio and from X-ray observations at galaxy cluster merger shocks may be due to an extension of DSA to superdiffusive transport \citep[][]{Zimbardo2017,Zimbardo2018}, which leads to superdiffusive shock acceleration \citep[][]{Perri2012}. Therefore, it is now timely to understand under what conditions the transport of energetic particles at shocks could be considered to be superdiffusive.

One of the main assumptions underlying DSA is that energetic particles undergo normal (i.e., Brownian) diffusion when scattering back and forth across the shock front. Brownian diffusion is characterized by Gaussian statistics. For an ensemble of particles undergoing normal diffusion, the mean square displacement for particle trajectories grows linearly in time, as $\langle\Delta s^2\rangle = 2 D t$, where $D$ identifies the so-called diffusion coefficient. 

If the time evolution of the MSD is nonlinear, for example, such that $\langle\Delta s^2\rangle = 2 D_\alpha t^\alpha$, then the system exhibits anomalous transport. The exponent $\alpha$ is known as the anomalous diffusion exponent and its value identifies the transport regime of the system. Anomalous transport has been observed in an enormous variety of systems, ranging from the foraging movements of spider monkeys \citep[][]{Fernandez2004} to stock options transactions in finance \citep[][]{Meerschaert2006}. It is possible to predict, upstream of shock waves, power-law density profiles for shock accelerated particles assuming for them an anomalous diffusion regime \citep[][]{Perri2007,Perri2008a}. In fact, an important feature of anomalous transport is that the systems are characterized by non-Gaussian statistics, which introduces probability distributions with power-law tails \citep[e.g.][]{Klafter1987,Metzler2000}. {In DSA, the time scale for acceleration of particles of given energy is dominated by the slowest portion of the shock acceleration cycle~\citep[e.g.,][]{Drury1983}. An interesting implication of anomalous, superdiffusive transport is that it may possibly lead to shorter acceleration times than those usually obtained for normal diffusive transport, thus helping to reach higher maximum energies than those predicted by DSA \citep[][]{Perri2015}.}

Simulations of particle acceleration at shocks have been used extensively to bridge theory and observations. Numerical and analytical solutions for the transport equation obtained from DSA have been studied under many approximations \citep[e.g.][]{Malkov1997,Kang1997,Caprioli2012}. A popular approach to address energisation efficiency and particle transport is the use of Monte Carlo simulations \citep[e.g.][]{Ellison1990,Jones1991,Baring1995,Wolff2015,Bykov2017}. A drawback of the usual Monte Carlo approach is the neglect of any self-consistent plasma effects (e.g., internal shock structure) which are believed to be important, particularly at low energies \citep[e.g.][]{Sundberg2016}. Efficient particle acceleration has been studied both in fully kinetic Particle-In-Cell (PIC) and hybrid (particle ions and fluid electrons) simulations \citep[e.g.][]{Amano2007,Amano2009,Riquelme2011,Giacalone1992,Giacalone2000,Caprioli2014a}. In the fully kinetic model, the spatial scales are resolved down to the Debye length, making it possible to address the complete picture of particle acceleration down to electron dynamics. However, fully kinetic PIC simulations are computationally intensive. On the other hand, in the hybrid model, the electrons are modelled as a fluid and just the ion kinetic scales are resolved, with the advantage that considerably larger domains and time scales can be simulated. 

Particle diffusion has also been studied in the framework of self-consistent plasma simulations. \citet[][]{Servidio2016}, for example, studied the transport of protons in different turbulent regions by means of hybrid simulations. The majority of plasma turbulence simulations are performed on a periodic domain and in the plasma rest frame, making it relatively easy to trace particle trajectories throughout the simulations and compute the MSD for them. However, for studying energetic particles in shock simulations the situation is more complex since only a small fraction of the particles are accelerated to high energies, they may exit the simulation domain because it is nonperiodic and there is no single rest frame due to the velocity change at the shock.

Although these particular features make it harder to track energetic particles in the simulations for sufficiently long times, early studies attempted to study particle transport at shocks using self-consistent simulations. For instance, \citet[][]{Kucharek2000} studied particle diffusion in turbulence generated by an ion-ion beam instability and invoked it as a model for the Earth's foreshock. \citet[][]{Scholer2000} used hybrid PIC simulations of turbulence to model the immediate downstream of a perpendicular shock (by using a particular initial condition, consisting of a homogeneous plasma core distribution and a non-gyrotropic ion population) and addressed the particle transport in such a system. \citet[][]{caprioli2014c} considered the shock transition layer of a hybrid PIC shock simulation at a single time, imposed boundary conditions on it and evolved test particles in the resulting electromagnetic fields to study diffusion across the shock layer. 

Our goal is to study the particle diffusion regimes at quasi-parallel shocks using self-consistent, hybrid PIC simulations, with the aim of gaining extra information that might advance our understanding of particle transport and so improve Monte Carlo modelling of shock acceleration.We focus on a set of shock parameters relevant for heliospheric shocks and intermediate particle energies. As we mentioned above, it is not trivial to follow many particles for long times in a shock simulation, and this has motivated us to employ a novel diagnostic, namely the time-averaged mean square displacement method (TAMSD). TAMSD relies on time-averaging rather than ensemble-averaging as in MSD, making it possible to extract single particle diffusion exponents from their trajectories \citep[][]{quian1991}. TAMSD has been used in a variety of systems where it can be difficult to obtain many particle trajectories to construct an ensemble average; this is the case for many biological systems, especially those for which \emph{in-vivo} tracking of particles is required \citep[see, for example][and references therein]{hafling2013,Weron2017,metzler2014}.

In this paper the trajectories for an ensemble of protons self-consistently interacting with a 1D, quasi-parallel shock are studied. Upstream and downstream diffusion properties are analysed separately, and all the computations are performed in the respective local rest frame. We obtain upstream and downstream diffusion exponents using both MSD and TAMSD. A thorough comparison between the two diagnostics is presented, in order to test the applicability of TAMSD. We have found the two techniques to be consistent, although both MSD and TAMSD have their own limitations. Upstream, strong evidence for superdiffusion is found. In the shock downstream, we observe a richer scenario, encompassing superdiffusion and subdiffusion, and, based on the TAMSD diagnostics we suggest that this is related to different subsets of particles co-existing in different diffusion regimes.
The paper is organised as follows: in Section~\ref{sec:method} we describe both the simulation method and the diagnostics employed. In Section~\ref{sec:results} results for upstream and downstream transport are described, and summarised in Section~\ref{sec:conclusions}. In Appendix \ref{appendix} we present some independent tests of the TAMSD diagnostic using Monte Carlo simulations.

\section{Methods}
\label{sec:method}
\subsection{Shock simulation}
\label{subsec:shocksim}
We simulate a shock using a hybrid PIC simulation. In the simulation, protons are modelled as macroparticles and advanced using the standard PIC method. The electrons, on the other hand, are modelled as a massless, charge-neutralizing fluid with an adiabatic equation of state. The hybrid code used throughout this work (HYPSI) is based on the CAM-CL algorithm \citep[][]{Matthews1994,Sundberg2016}. 

The shock is initiated by the injection method, in which the plasma flows in the $x$-direction with a defined (super-Alfv\'enic) velocity $V_\mathrm{in}$. The right-hand boundary of the simulation domain acts as a reflecting wall, and at the left-hand boundary plasma is continuously injected. The simulation is periodic in the $y$ direction. A shock is created as a consequence of reflection at the wall, and it propagates in the negative $x$-direction. In the simulation frame, the upstream flow is along the shock normal.

In the hybrid simulations distances are normalised to the ion inertial length $d_i \equiv c/\omega_{pi}$, times to the inverse cyclotron frequency ${\Omega_{ci}}^{-1}$, velocity to the Alfv\'en speed $v_A$ (all referred to the upstream state), and the magnetic field and density to their upstream values, $B_0$ and $n_0$, respectively. The angle between the shock normal and the upstream magnetic field, $\theta_{Bn}$ is 15$^\circ$, with the upstream magnetic field  in the $x$-$y$ plane. For the upstream flow velocity, a value of  $V_\mathrm{in} = 6 v_A$  has been chosen, and the resulting Alfv\'enic Mach number of the shock is approximately $M_A = 8$. The upstream ion distribution function is an isotropic Maxwellian and the ion $\beta_i$ is 0.5.  The simulation $x-y$ domain  is 3000 $\times$ 10 $d_i$, and hence we describe the shock as quasi-1D. The spatial resolution used is $\Delta x$ = $\Delta y$ = 0.5 $d_i$. The final time for the simulation is 750 $\Omega_{ci}^{-1}$, the time step for particle {(ion)} advance is $\Delta t_{}$  = 0.005 $\Omega_{ci}^{-1}$. Substepping is used for the magnetic field advance, with an effective time step of $\Delta t_{B} = \Delta t_{}/10$. A small, nonzero  resistivity is introduced in the magnetic induction equation. The value of the resistivity is chosen so that there are not excessive fluctuations at the grid scale. The number of particles per cell used is always greater than 300 (upstream), in order to keep the statistical noise characteristic of PIC simulations to a reasonable level.

The number of particles in the simulation is about 10$^7$, and we store the trajectories of about 5 $\times$ 10$^5$ particles. The time cadence for the particle trajectory outputs is chosen such that the time resolution obtained in the particle dataset is $\Delta t$ = 0.1 $\Omega_{ci}^{-1}$. The protons are tracked in time starting at 250 $\Omega_{ci}^{-1}$,  when the shock is already well-developed, with an upstream region with developed fluctuations and a well-defined downstream region separated from the right hand wall.

\begin{figure*}
    \includegraphics[width=\textwidth]{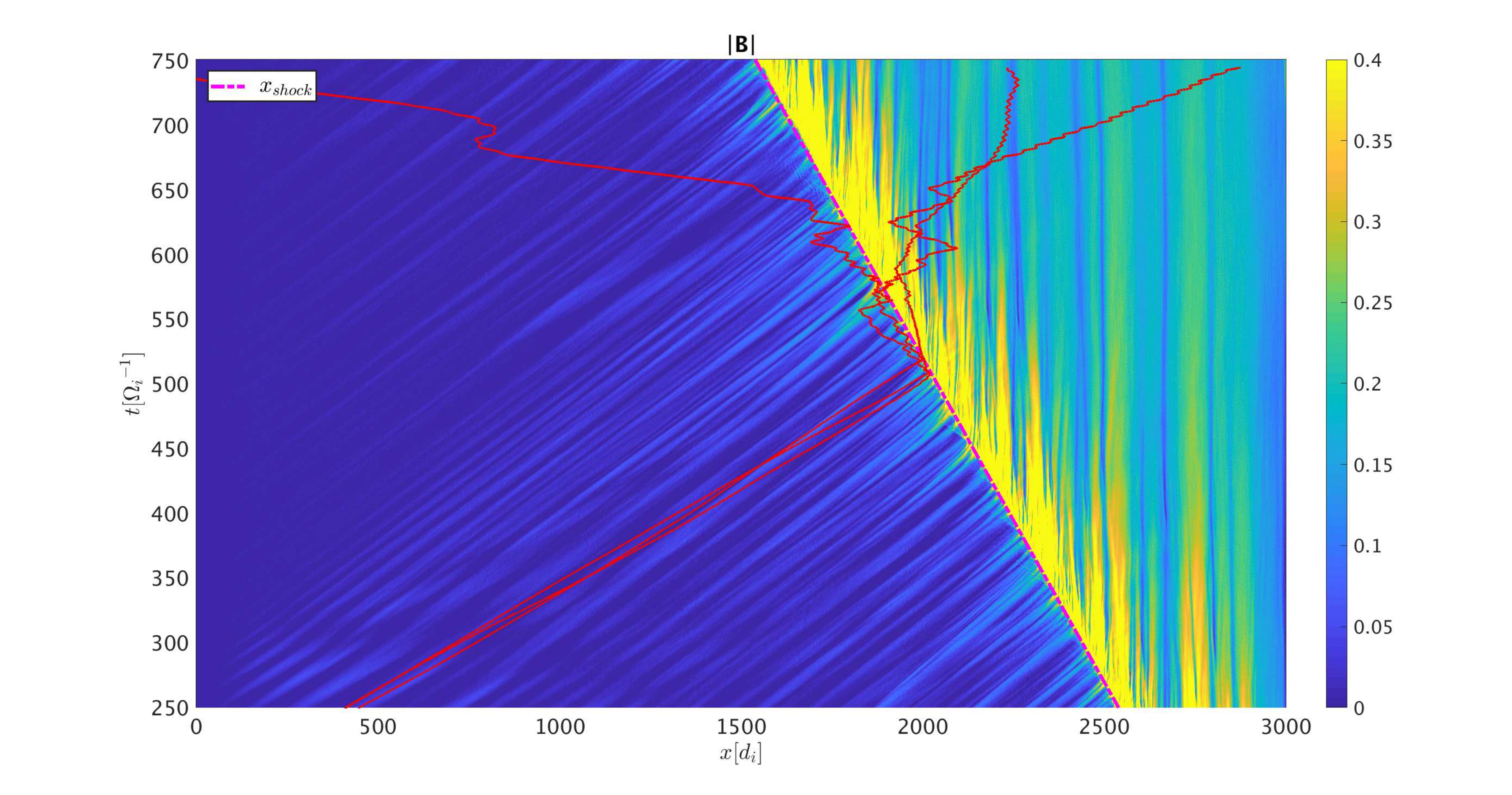}
    \caption{Total magnetic field as a function of space ($x$-axis) and time ($y$-axis). The colorbar shows the total magnetic field magnitude in logarithmic scale.  The magenta dotted-dashed line identifies the shock nominal position. Red lines represent three examples of protons interacting with the shock. }
    \label{fig:shk_overview}
\end{figure*}

An overview of the shock simulation is presented in Figure~\ref{fig:shk_overview}, which shows the time evolution of $|B|$ as a function of $x$ position. Three typical trajectories for protons interacting with the shock are also shown (red solid lines). Throughout the simulation, there are upstream waves self-consistently induced by the reflected ion population, that have been extensively studied in previous literature \citep[e.g.,][]{Sundberg2016}. The downstream region exhibits strong fluctuations at several wavelengths, and some of them are relatively coherent in time, an effect related to the quasi-1D nature of the simulation.

As can be seen in Figure~\ref{fig:shk_overview}, the upstream waves are artificially truncated. This effect, particularly evident at later times, is due to the finite size of the simulation domain, and is known to affect shock simulations. It can be commented that this is not just a spatial truncation, but also an amplitude truncation, because the extent of the current and the growth (advection) time is truncated. Also, if ions with a given energy escape before properly producing waves, it is likely that particles with the same energy will see a non-fully-developed set of fluctuations, and this could affect their transport properties. However, the results presented here have been confirmed using larger simulations (in spatial extent and time duration), in order to ensure their robustness.

We define the nominal shock position $x_\mathrm{sh}$ in the simulation frame as the position at which the magnitude of the magnetic field exceeds the upstream value by at least a factor of 3. The nominal shock position is then fitted in time to define an average shock frame (dashed magenta line in Figure~\ref{fig:shk_overview}). The nominal shock position is used to systematically distinguish between upstream and downstream of the shock. The shock speed in the simulation frame (which is also the average downstream frame) is around 2 $v_A$.

\begin{figure*}
    \includegraphics[width=\textwidth]{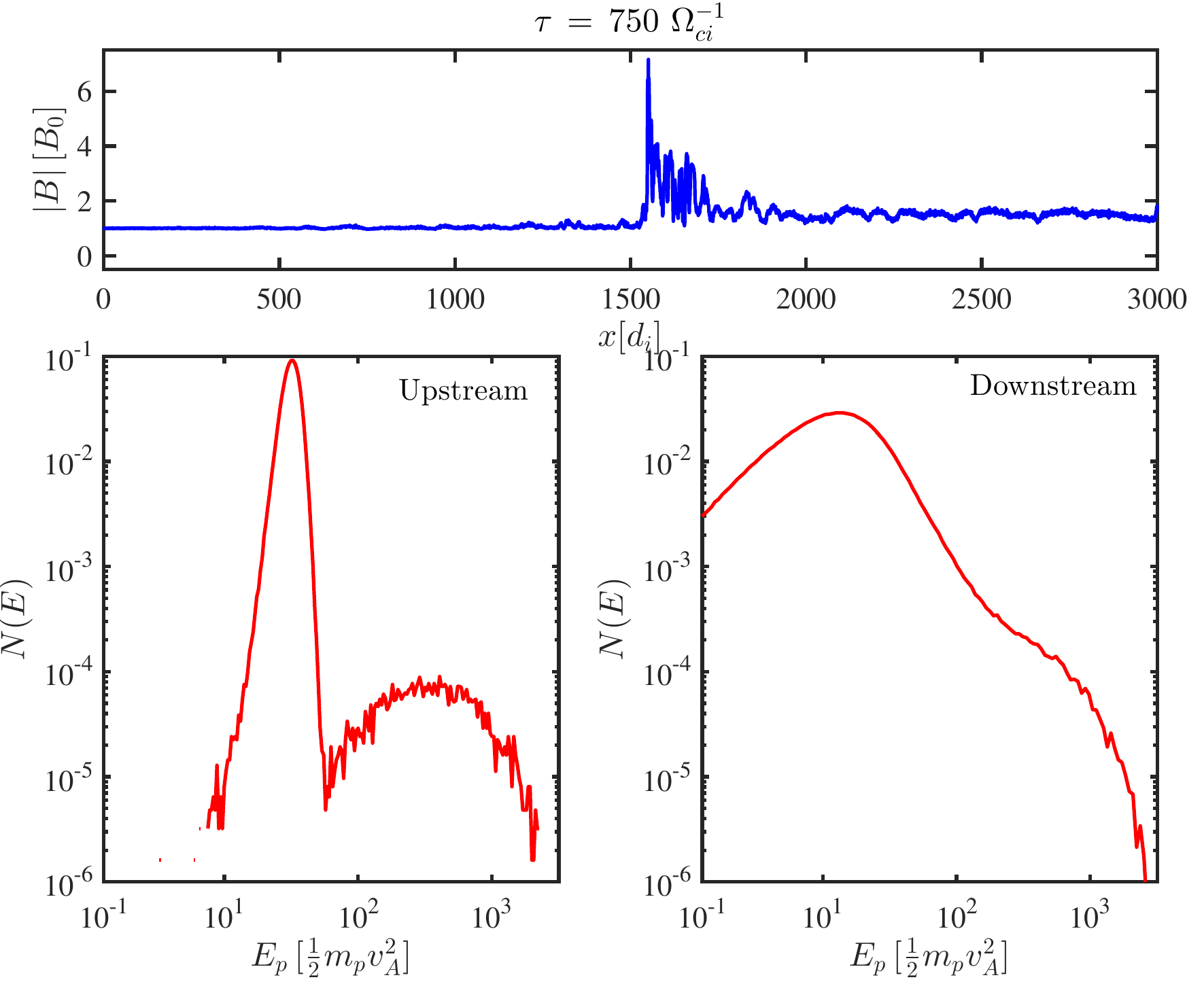}
    \caption{Top: Magnetic field magnitude (averaged over $y$) at the final time of the simulation.  Bottom: Upstream (left panel) and downstream (right panel) particle energy spectra calculated in the upstream and downstream rest frames, respectively.}
    \label{fig:spectra}
\end{figure*}

Figure~\ref{fig:spectra} shows the magnetic field magnitude at the final time of the simulation (top panel). The upstream region is defined as the simulation domain portion with $x < x_\mathrm{sh} - 25 \, d_i$, and it reduces in size as the simulation evolves. The downstream region is defined as the simulation domain portion with $x > x_\mathrm{sh} + 80 \, d_i$, which grows with the simulation time. The shock transition region, where there is the main change of flow velocity, is not considered in this work, because there is no well-defined, single flow frame for the analysis of particle diffusion regime.

Energy spectra for upstream and downstream populations are shown in the bottom panels of Figure~\ref{fig:spectra}. In both spectra, high energy, non-Maxwellian tails are present, extending up to energies corresponding to velocities of 80 $v_A$. Although this result is well known and has been widely investigated in previous literature \citep[e.g.,][]{Scholer1990, Giacalone2004, Sugiyama2011a,Gargate2012, Caprioli2014a}, it is important to underline that hybrid simulations of quasi-parallel shocks show significant particle acceleration, with energy gains of up to two orders of magnitude, even though the typical simulation times are shorter than the ones used in Monte Carlo simulations and other frameworks. The focus of this work is to find out more about the transport regime for particles in the high energy tail of the distribution, that is relevant to study the initial energisation of ions through their interaction with a quasi-parallel shock.

A pre-selection on the dataset of tracked particles is performed: we track particles having final energies above a threshold corresponding to a speed of 10 $v_A$ in the downstream frame. Fragments of upstream trajectories tracked before the first shock encounter are ignored in order to exclude the initial convection of the particle in the upstream flow towards the shock. This has the effect that the initial position for all trajectory fragments will be at the origin in the shock frame.

\begin{figure*}
    \includegraphics[width=\textwidth]{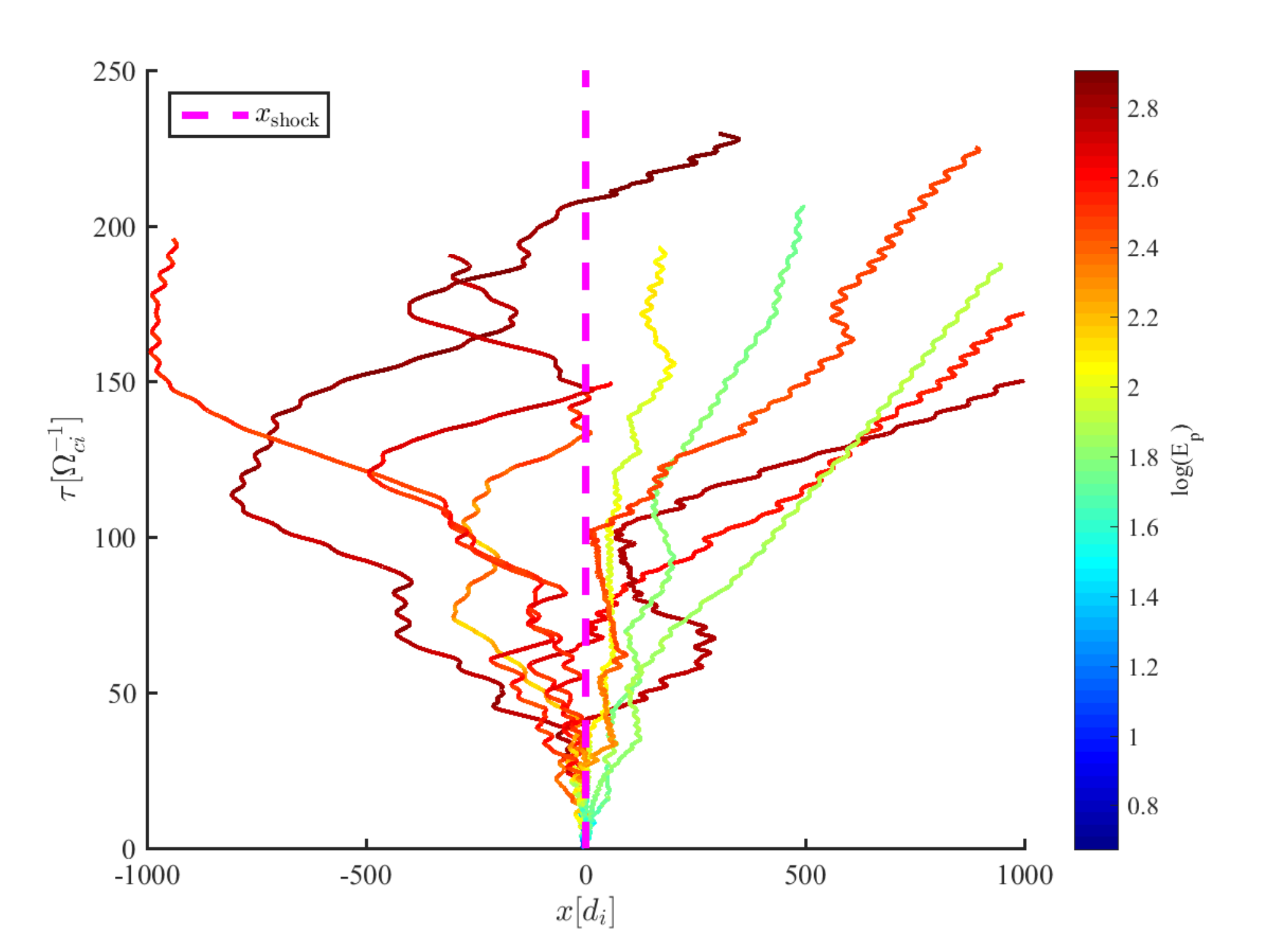}
    \caption{Typical particle trajectories extracted from the sample, in the shock frame. The colour bar shows the particle energy in logarithmic scale and in units of $\frac{1}{2}m_p v_A^2$. The dashed magenta line represents the nominal shock position.}
    \label{fig:smpl_oview}
\end{figure*}

Figure~\ref{fig:smpl_oview} shows some examples of particle trajectories in the shock frame. Even at a first glance, it is possible to note that trajectories with completely different dynamics coexist in the simulation: some particles are transmitted downstream and moving away from the shock, other particles seem to move at the same speed of the shock in the region immediately downstream of it, some other ones (the most energetic) are reflected upstream and have multiple encounters with the shock transition.

Diffusion diagnostics can be biased when performed in a drifting frame. Hence, we will treat the upstream and downstream regions of the simulation as independent, with all the calculations performed in the local rest frame. However, the particles are free to travel in between regions so that a particle crossing from one region to another is treated as a new particle in the new region. This ultimately enables us to deal with two different trajectory datasets (or, more accurately, with two different datasets of trajectory fragments).

It is likely that in limiting the study of the particle trajectories to one side only, either upstream or downstream, a bias is introduced in the statistics. We note, however, that for quantities like the cycle time for shock acceleration what matters are the transport regimes and the sum of times spent on each side of the shock before re-crossing the shock \citep{Drury1983,Perri2012_2,Zimbardo2013}. 

\subsection{Anomalous diffusion diagnostics}
\label{subsec:anomalous_diagnostics}
The particle mean square displacement (MSD) is the most common tool to address the particle diffusion regime in a system. For one-dimensional motion, it is defined as 
\begin{equation}
    \label{eq:eq1}
    \langle \Delta s^2 (t) \rangle  = \frac{1}{N_p} \sum_{n=0}^{N_p} [x_n(t) - x_n(0)]^2 ,
\end{equation}
where $N_p$ is the total number of particles in the ensemble, $x_n(t)$ and $x_n(0)$ are the $n$-th particle positions at time $t$ and at the initial time $t=0$, respectively. From now on, we will use the symbols $\langle a\rangle$ to denote ensemble averages, and $\overline{a}$ to indicate time averages. Note that the MSD is an ensemble average, telling us information about the overall behaviour of the system. Obviously, the accuracy of MSD has a strong dependence on the number of particles used to build the ensemble. Also, in a system with open boundaries, $N_p$ is not necessarily constant, and reduces in time.

In the case of anomalous diffusion the MSD is known to scale with time as
\begin{equation}
    \label{eq:eq2}
    \langle \Delta s^2 (t)\rangle = D_\alpha (E) \,\,t^\alpha ,
\end{equation}
where $D_\alpha(E)$ is the anomalous diffusion coefficient, {which, in general, depends on the particle energy}. The anomalous diffusion exponent $\alpha$ characterizes the diffusion regime, such that when $\alpha$ is equal to 1 Brownian diffusion is at play, whereas if $\alpha < 1$ or $\alpha > 1$ the system is subdiffusive or superdiffusive, respectively. It is possible, given an ensemble of particles for which the diffusion regime is not known, to infer a diffusion exponent, by calculating  $\langle \Delta s^2 (t) \rangle$  for the ensemble of particles, and then $\alpha$ can be estimated from a linear fit in log-log space.

Often, datasets of trajectories are not ideal for calculating the MSD, since difficulties can arise when dealing with poor particle statistics, leading to noisy ensemble averages. Furthermore, if the system is heterogeneous (e.g., different particle species present, non-uniform background environment), the MSD cannot account for this and simply gives an ensemble-averaged perspective on the overall system diffusion regime.

The time-averaged mean square displacement (TAMSD) method tries to address such problems by using single-particle trajectories and it is commonly used in many areas of biophysics \citep[e.g.][]{Golding2004,Wong2004,Saxton2012}. In TAMSD, the ensemble average is replaced by an averaging time window:
\begin{equation}
    \label{eq:eq3}
    \overline{\Delta s^2} (\tau) = \frac{1}{T_f - \tau} \sum_{m=1}^{T_f -\tau} [x(m+\tau) - x(m)]^2 ,
\end{equation}
where $\tau$ is the averaging window, $T_f$ is the final time for which the particle is tracked and $x$ is the particle position in time. All times are considered to be discretised with uniform sampling. This leads to an estimator of the anomalous exponent based on single particle trajectories if $\overline{\Delta s^2}$ scales with time in the same way described by Equation~\ref{eq:eq2}. In the ideal case, for a dataset with infinite time records and for Brownian motion, TAMSD and MSD would produce the same result because of the central limit theorem.

The general procedure to apply TAMSD to a sample of trajectories is to fit an anomalous exponent $\alpha$ to each particle trajectory (as for MSD), and then take an average of the obtained exponents $\langle\alpha\rangle$. This is known to have some problems that often lead to an underestimation of $\langle\alpha\rangle$, deriving from the fact that new free parameters are introduced in the diagnostic \citep[see, for example,][]{Kepten2015,Burnecki2015}. For example, a time interval needs to be chosen for the fitting of the exponents, and this is nontrivial particularly when long-memory processes are present in the system. However, a set of particles interacting with a shock is intrinsically heterogeneous, and we suggest TAMSD as a tool to identify different ensembles of particles with different diffusion properties.

To have direct comparison between MSD and TAMSD, it is possible to consider an ensemble average of the sample's TAMSD \citep[][]{Kepten2013}. We employ this diagnostic, called ensemble-averaged time-averaged mean square displacement (EA TAMSD), to the particle trajectories obtained in the shock simulation. The EA TAMSD is defined by 
\begin{equation}
    \label{eq:eq4}
    \langle\overline{\Delta s^2}  (\tau)\rangle = \frac{1}{N_p} \sum_{p=1}^{N_p} \frac{1}{T_f - \tau} \sum_{m=1}^{T_f -\tau} [x_p(m+\tau) - x_p(m)]^2 ,
\end{equation}
that is the average of the TAMSD in Equation \ref{eq:eq3} over the ensemble of tracked particles. As we will show in the results section, due to its double-averaging feature, the EA TAMSD is not only a way to validate the results found using MSD, but is also a very efficient diagnostic to investigate the overall (i.e., ensemble averaged) properties of the sample. Comparing MSD and EA TAMSD is also important because previous studies have shown that  MSD and EA TAMSD may differ when there is an ergodicity breaking in the sample \citep[][]{metzler2014}.

Even though this work is concerned with computing diffusion exponents from particle trajectories obtained from hybrid PIC shock simulations, MSD, TAMSD and EA TAMSD diagnostics have also been tested on a  Monte Carlo model, to ensure the robustness of the inferences made for the PIC simulations. Two examples of such tests are given in Appendix ~\ref{appendix}.

\section{Results}
\label{sec:results}
\subsection{Upstream particle transport}
\label{subsec:upstream}
Results regarding the diffusion regimes for particles upstream of the shock are presented in this section. All the calculations shown here are done in the upstream rest frame. The particle sample considered consists only of particles initially reflected at the shock, as the trajectory fragments of particles before the first interaction with the shock are neglected. Throughout this analysis, particles are treated regardless of their energies even though the particle diffusion coefficient is known to be energy dependent, and this may affect the estimation of transport properties. However, this choice is motivated by the fact that the discussion below is focused on investigating the particle transport regime using different diagnostics (MSD, TAMSD, EA TAMSD) rather than to provide a discussion about particles' diffusion coefficients. However, an analysis of particle MSD for different energies is reported in Section~\ref{subsec:energy}.
It is worth noting that the upstream sample is the poorest from the statistical point of view, as the particles interacting with the shock have a relatively high probability to be transmitted downstream. On the other hand, the particles in the upstream region are likely to be efficiently accelerated, as they can gain energy through the multiple-shock encounter scenario discussed above and seen in the trajectories of Figure~\ref{fig:smpl_oview}.

\begin{figure*}
    \includegraphics[width=\textwidth]{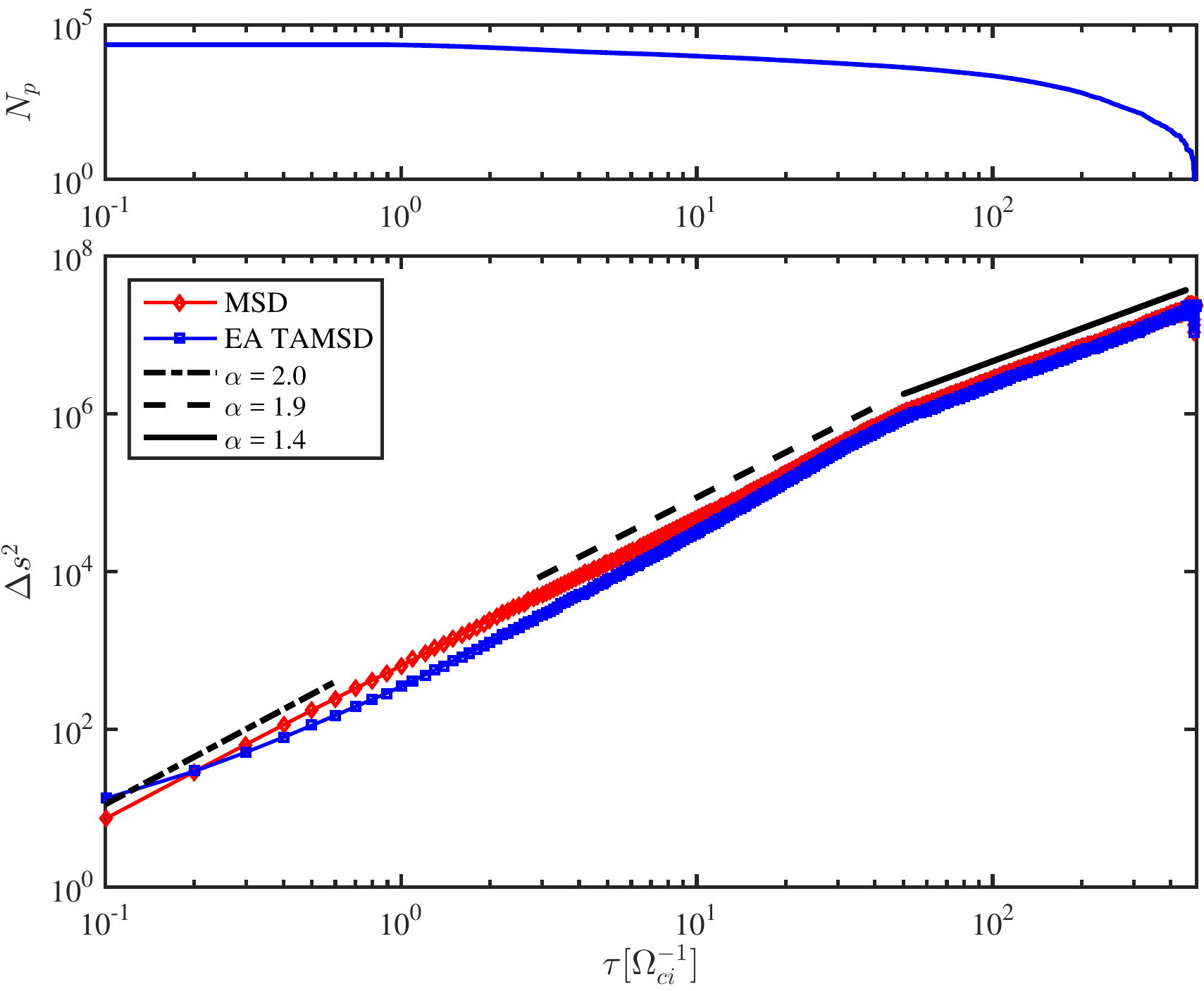}
    \caption{Top: Number of particles in the upstream region as a function of time. Bottom: MSD and EA TAMSD for upstream sample (red and blue lines, respectively). The black lines show the diffusion exponents $\alpha$ obtained by fitting the MSD in the indicated ranges.}
    \label{fig:up_msd}
\end{figure*}

MSD and EA TAMSD are shown in Figure~\ref{fig:up_msd}, which gives an overall view of the ensemble behaviour upstream of the shock.
Firstly, we observe that the overall behaviour of the two diagnostics is remarkably similar, validating the use of EA TAMSD. Note that the time $\tau$ shown when comparing MSD and EA TAMSD in Figure ~\ref{fig:up_msd} has two different meanings: for the MSD it is the proper simulation time (see Equation~\ref{eq:eq1}) and for the EA TAMSD it is the length of the averaging time window, defined in Equation~\ref{eq:eq3}. 

At extremely short times (less than one $\Omega_{ci}^{-1}$), a ballistic regime, with the diffusion exponent $\alpha$ =2 is identified in the MSD. This is expected since at short times scattering is still not taking place and, in principle, both diffusion and superdiffusion should be studied over sufficiently long times (i.e., for times longer than a few interaction times). Since the ballistic regime lasts for a very short time, the EA TAMSD does not capture well this ballistic transient, and this is related to the time average operation included in TAMSD.

At moderate times (from 6 to 45 $\Omega_{ci}^{-1}$), there is a slight change of behaviour, and the MSD show a quasi-ballistic trend. A fit to the displacement in time gives values for the anomalous diffusion exponent of 1.86 and 1.93 when calculated from MSD and EA TAMSD, respectively. These values are both consistent with superdiffusion. At later times (from 50 to 450 $\Omega_{ci}^{-1}$), another behaviour can be seen where MSD and EA TAMSD have exactly the same slope, and enter another superdiffusive regime, with an anomalous diffusion exponent $\alpha$ of 1.4. The reason for this change of slope may be that the most superdiffusive particles (i.e., the ones undergoing a very small number of scattering episodes) leave the box on short timescales.

It is important to underline that, for the upstream sample, EA TAMSD and MSD have a very similar behaviour and are almost indistinguishable for more than a decade. On one hand, this is a good test for the TAMSD diagnostic, which is a novel technique for particle simulations in collisionless plasma. On the other hand, this result is a validation for what was found using MSD.

TAMSD has the advantage that it gives an insight into the diffusive behaviour of single particles, in particular, diffusion exponents for each single particle trajectory.
\begin{figure*}
    \includegraphics[width=\textwidth]{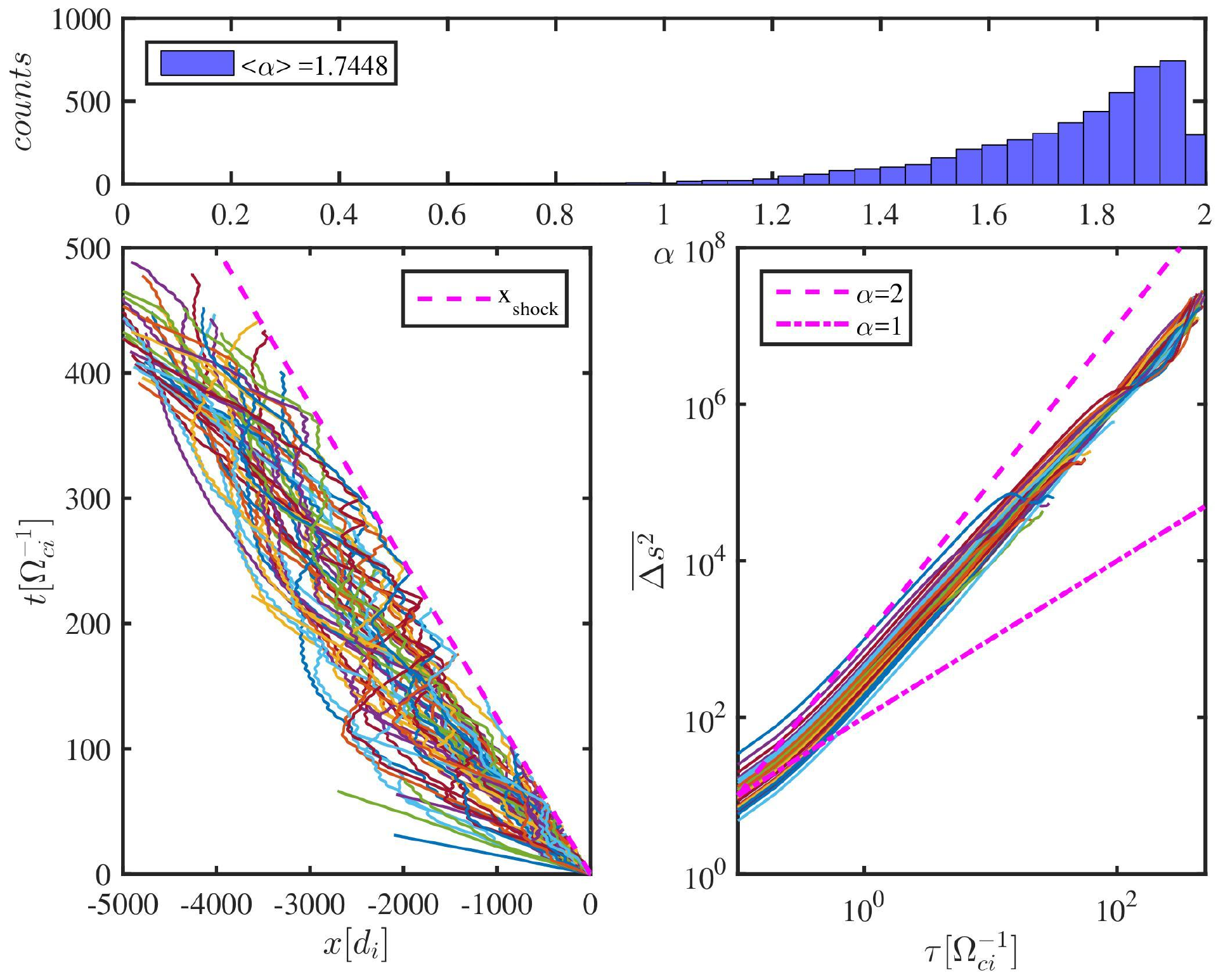}
    \caption{Top: Histogram of single particle diffusion exponents obtained from TAMSD for the upstream region. Bottom: Examples of a subset of typical upstream particle trajectories (left) and their single particle TAMSDs (right). Lines corresponding to $\alpha = 1$ and $\alpha =2$ are also plotted.}
    \label{fig:up_hist_trj}
\end{figure*}
Figure~\ref{fig:up_hist_trj} shows an example of how single particle TAMSD can be used for the upstream particles. A subset of around 4000 trajectory fragments of particles surviving for at least 200 $\Omega_{ci}^{-1}$ in the upstream region has been selected. Single particle TAMSDs have been calculated using Equation \ref{eq:eq3} and then single particle diffusion exponents have been fitted to obtain a distribution of single particle diffusion exponents $\alpha$.  The histogram in Figure~\ref{fig:up_hist_trj} (top) shows the values of $\alpha$ obtained with TAMSD on these trajectory fragments. The fit excludes the last few points of the single particle TAMSD in order to reduce the statistical noise present when using large averaging windows (see TAMSD fluctuations at large $\tau$ in Figures~\ref{fig:up_hist_trj} \&~\ref{fig:dw_tamsd}) and to obtain a robust estimation of the values of $\alpha$.

The fluctuations that TAMSD shows at long time scales are due to poor statistics in the trajectory fragments. From Equation~\ref{eq:eq3}, for larger times fewer points contribute to the time averaging window, so if a particle returns close to its initial position the value of $\overline{\Delta s^2}$ drops drastically, inducing the fluctuation seen in the single particle TAMSD.

The mean value for the single particle $\alpha$ distribution  shown in Figure \ref{fig:up_hist_trj} is approximately 1.75, i.e., between the transient with very strong superdiffusion (almost ballistic) and the long term superdiffusive behaviour of 1.4 found using MSD and EA TAMSD and shown in Figure~\ref{fig:up_msd}. This value for $\langle\alpha\rangle$ is still consistent with superdiffusion. We note that a transition between different anomalous transport regimes is also found in laboratory plasmas \citep[][]{Bovet2015}. It is also worth noting that every single particle $\alpha$ value is consistent with superdiffusion.

%% The upstream particle trajectories examples are also consistent with what
%% we found for the $\alpha$s: they are in fact consistent with L\'evy walks,
%% i.e.,  alternating very long jumps with scattering episodes.

\subsection{Downstream particle transport}
\label{subsec:downstream}
In this section we present results obtained for the region downstream of the shock. All the calculations are done in the downstream rest frame. The downstream region is very different from that upstream since the plasma has been thermalised by the shock and the level of fluctuations is much higher. From the point of view of particle transport, particles with very different histories are found downstream. Some particles are transmitted and moving away from the shock, others are returning upstream (gaining energy in the process), and some others stay trapped in the strong magnetic field fluctuations present immediately after the shock transition. Some traces of this intrinsic heterogeneity are expected to be found when addressing the diffusion regime.

{With the same motivation discussed for the upstream region, particles are treated regardless of their energy. A discussion about the downstream MSD behaviour at different energies is reported in Section~\ref{subsec:energy}}. 

An ensemble of downstream particle trajectory fragments is built in the same fashion as described earlier. A lower boundary is set just after the shock transition zone (80 $d_i$ downstream of the nominal shock transition). Another boundary is placed close to the reflecting wall (20 $d_i$ in front) to remove any artificial scattering that the wall might produce, and when a particle crosses this ``wall boundary'' it is discarded. Every particle that crosses the left boundary to enter the downstream region is treated as a new particle trajectory fragment, and particles escaping the left boundary are discarded and tracked as new particles in the transition zone.
\begin{figure*}
    \includegraphics[width=\textwidth]{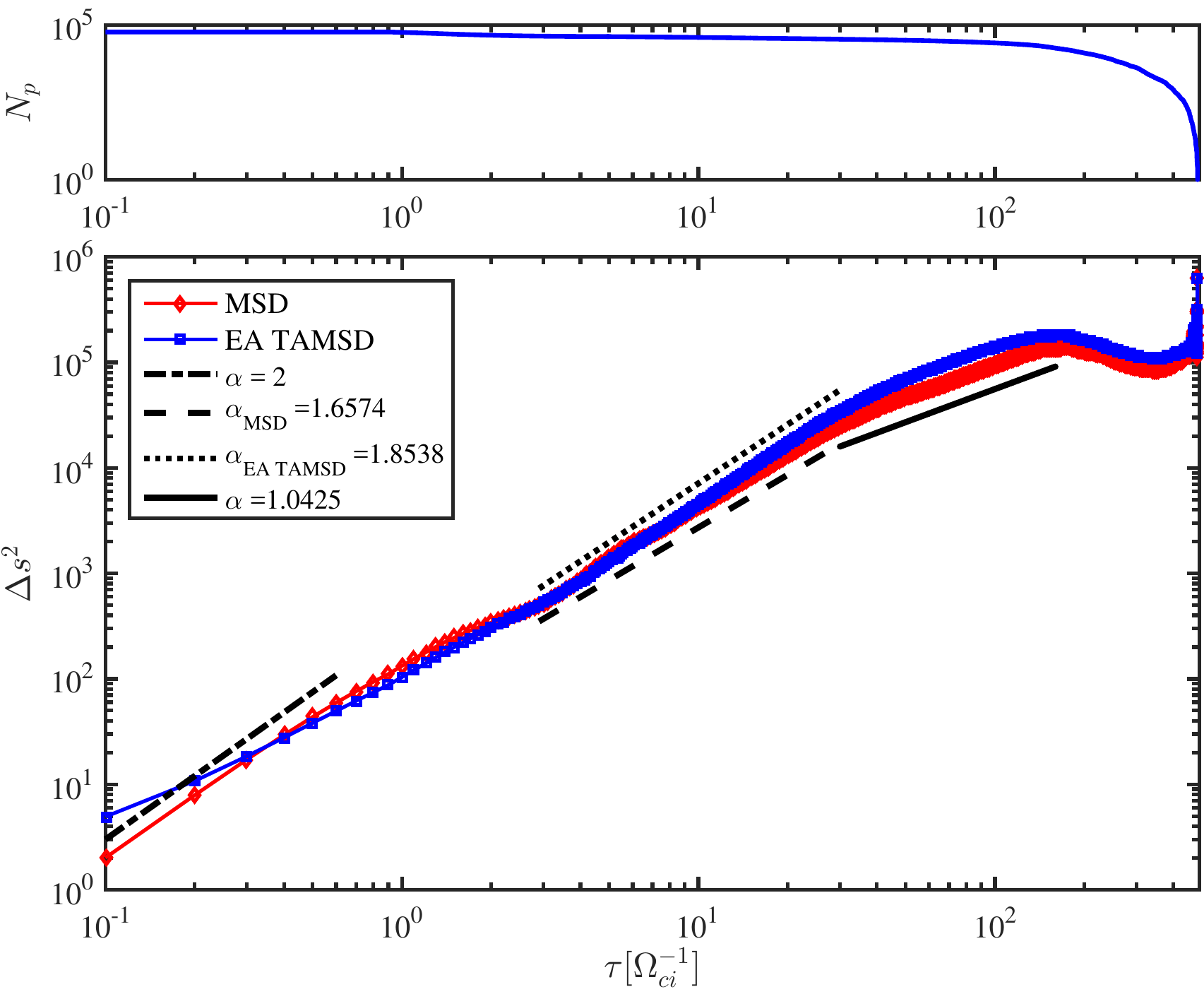}
    \caption{Top: Number of particles in the downstream region as a function of time. Bottom: MSD and EA TAMSD for the downstream particle sample (red and blue lines, respectively). The black lines show the diffusion exponents $\alpha$ obtained by fitting the MSD in the indicated ranges.}
    \label{fig:dw_msd}
\end{figure*}
Figure~\ref{fig:dw_msd} shows a comparison between MSD and EA TAMSD for the downstream sample. The scenario obtained is clearly different from the upstream case. At very short times, the ballistic behaviour is recovered with MSD. Once again, the EA TAMSD does not capture well this regime because of the time average included in the diagnostic. At intermediate times (from 3 to 30 $\Omega_{ci}^{-1}$), a temporarily superdiffusive regime is observed once again. The behaviour at intermediate times is different from the one found in the upstream particle sample. First of all, the diffusion exponents found with MSD and EA TAMSD, 1.65 and 1.85 respectively, are smaller than the typical value of 1.9 found upstream. This result can be interpreted considering that scattering for downstream particles starts to happen, statistically, at smaller time scales than the upstream case, and this is due to the higher level of fluctuations downstream of the shock. It is important to note that, for intermediate times, the diffusion exponents obtained with MSD and EA TAMSD differ by 0.2, which may be due to the heterogeneity of the sample and the process of ensemble averaging, as discussed later. 

For longer times (30 to 150 $\Omega_{ci}^{-1}$), both diagnostics have the same slope, with an $\alpha$ very close to unity, indicating normal diffusion. The recovery of normal diffusion yields to two important considerations. On one hand, we find a diffusion regime that is compatible with the theoretical assumptions of DSA, but on the other hand, the result must be interpreted keeping in mind that the ensemble average involved in both the diagnostics averages out the behaviour of single particles or smaller populations of particles that may be in different transport regimes.

With single particle TAMSD, we can identify subsets of particles having different behaviours. Before discussing single particle TAMSD, it is worth  noting that, for times larger than 150 $\Omega_{ci}^{-1}$,  $\Delta s^2$ exhibits a large scale oscillation. This is an effect due to the increase of the finite size of the downstream region, confirmed by performing different simulations with different box sizes.     
\begin{figure*}
    \includegraphics[width=\textwidth]{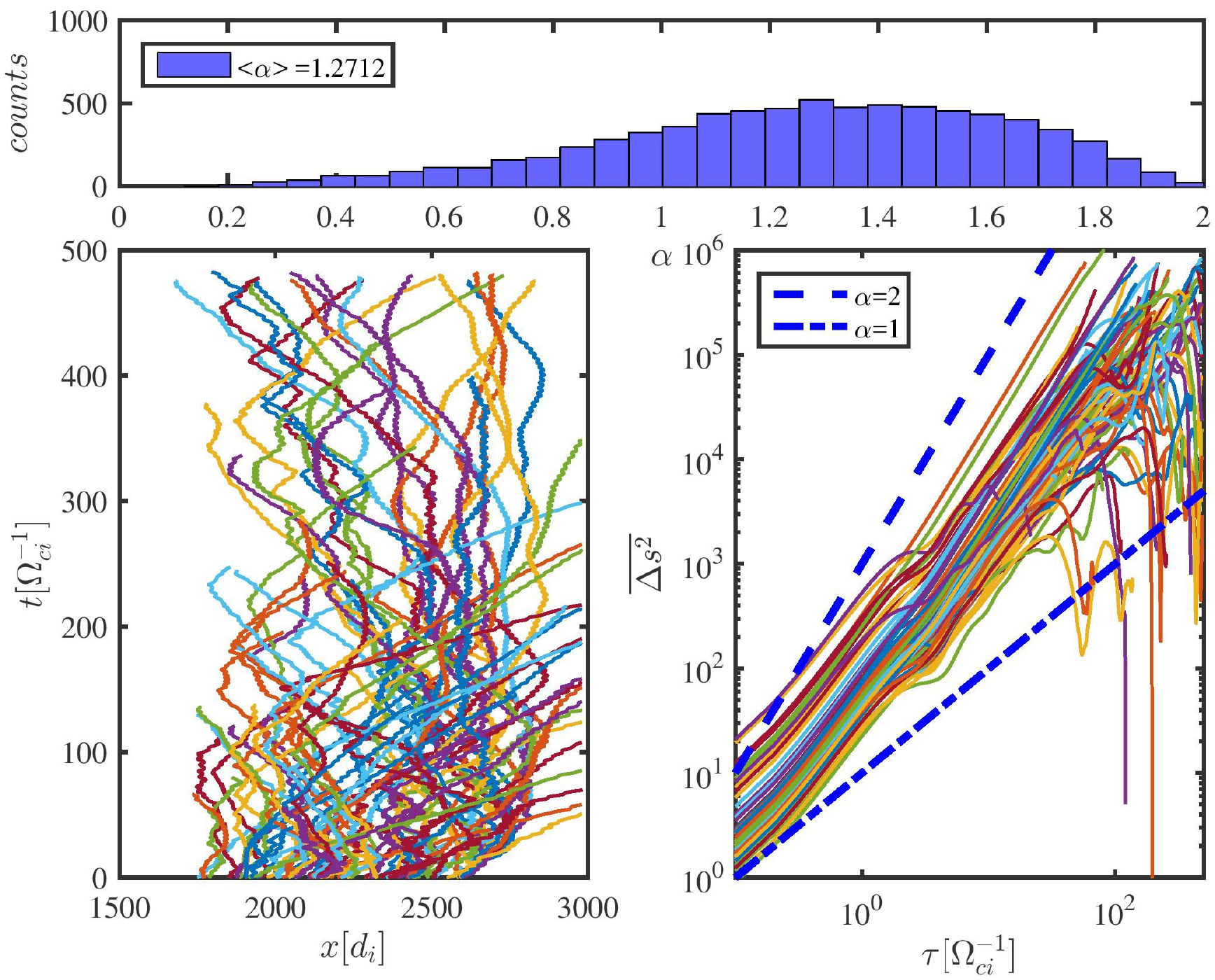}
    \caption{Top: Histogram of single particle diffusion exponents obtained from TAMSD for the downstream region. Bottom: Examples of a subset of typical downstream particle trajectories (left) and single particle TAMSDs (right).}
    \label{fig:dw_tamsd}
\end{figure*}
Results for single particle TAMSD are shown in Figure~\ref{fig:dw_tamsd}. Again, a subset of particles that survive for at least 200 $\Omega_{ci}^{-1}$ has been selected to perform this diagnostic. The single particle diffusion exponents have been fitted excluding the last few points of $\overline{\Delta s^2}$, where TAMSD become noisy (as for the upstream region). The histogram representing the $\alpha$ values obtained for the particle ensemble has a mean value of 1.27, representing superdiffusion. Although this value of $\langle\alpha\rangle$ is smaller than that found upstream (c.f. $\langle\alpha\rangle = 1.74$), it is still indicative of superdiffusion with an exponent consistent with those obtained from data analysis \citep[e.g.][]{Perri2015b,Zimbardo2018}. Like the upstream case, the $\langle\alpha\rangle$ obtained with TAMSD is in between the two regimes found with MSD and EA TAMSD (see Figure~\ref{fig:dw_msd}). 

When confronting the results obtained with different diagnostics, it is useful to note that EA TAMSD (defined in Equation \ref{eq:eq4}) leads to an estimation of $\alpha$ over the ensemble, that may depend on time, and is robust against the noise of MSD. TAMSD, on the other hand, leads to an estimation of single particle $\alpha$ values without $\alpha$ varying in time, so it introduces the possibility of examining the heterogeneity of the ensemble. To this extent, when analysing the downstream particle transport properties, the key result is not the mean value of the TAMSD histogram, but rather its large spread over values covering the whole domain of possible anomalous transport exponents, that demonstrates that particles with completely different diffusion regimes coexist in the downstream region.

It can be seen in the bottom-right plot of Figure~\ref{fig:dw_tamsd} that single particle TAMSD have a remarkable spread, much larger than those observed in the upstream samples, ranging from subdiffusive regimes ($\alpha < 1$) to superdiffusive ($\alpha > 1$). We suggest that subdiffusive diffusion exponents represent particles trapped in the strong magnetic fluctuations present downstream of the shock (and in particular immediately after the shock transition). Superdiffusive single particle anomalous exponents can, on the other hand, correspond to particles transmitted downstream which move away from the shock without undergoing much scattering.

\subsection{TAMSD advantages and limitations}
\label{subsec:tamsd}

\begin{figure*}
    \includegraphics[width=\textwidth]{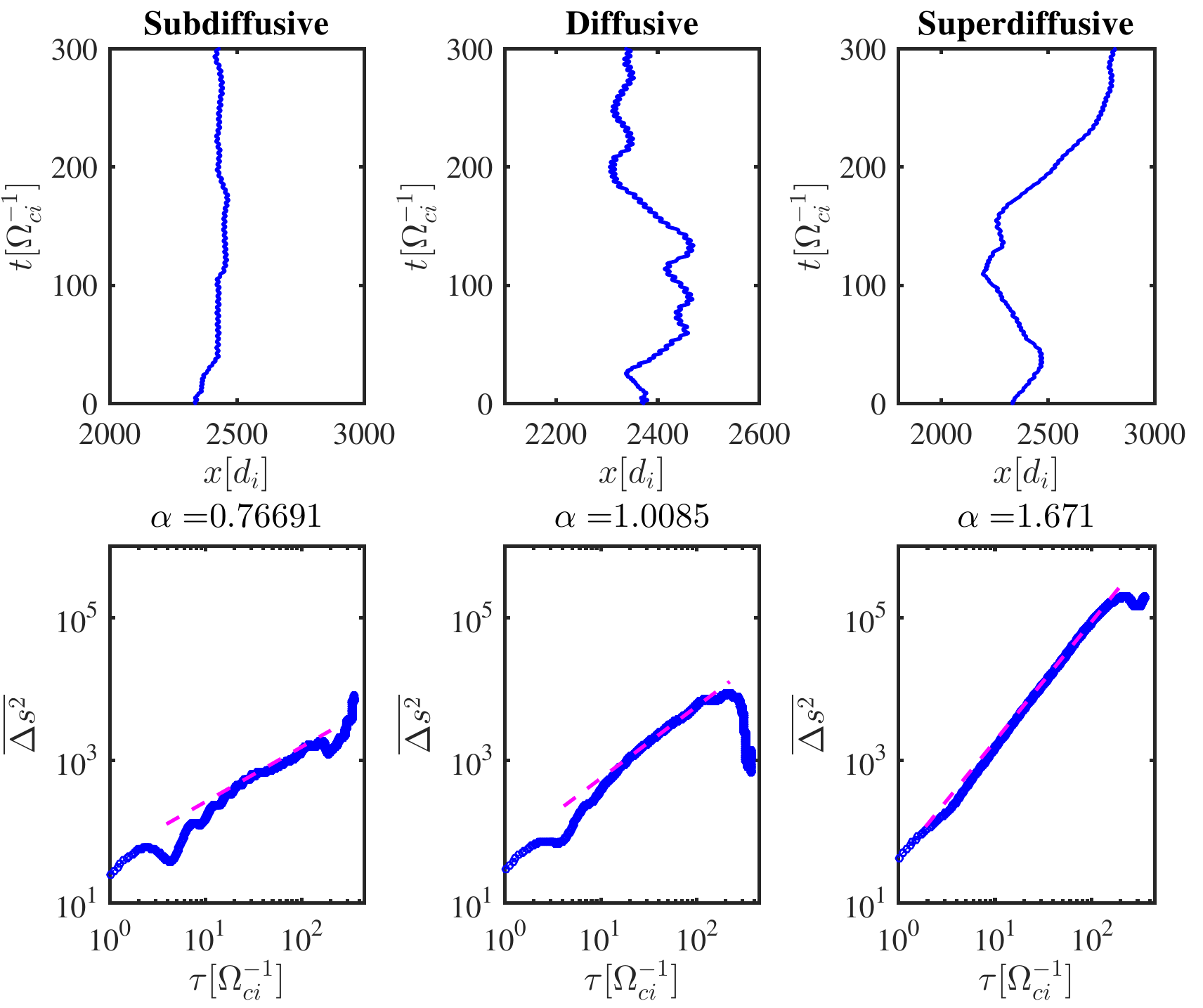}
    \caption{Top: Three particle trajectory examples. Bottom: Single particles TAMSDs (blue lines) and their fits (dashed magenta lines).}
    \label{fig:dw_traj}
\end{figure*}

To demonstrate our interpretation of the many possible transport realizations seen in the downstream region, we show in Figure~\ref{fig:dw_traj} three examples of particles in different transport regimes.  
 On the left, a particle trapped downstream is shown. The $\alpha$ value obtained with single particle TAMSD is 0.7, indicating a subdiffusive behaviour. The same analysis is carried out in the middle plot, and gives a normal diffusive value of $\alpha$, confirmed from the particle trajectory observed, that undergoes many scattering episodes but still samples more space than for the particle shown in the left panel. Finally, a particle with superdiffusive $\alpha$ value is shown (right panel), and it is an example from the population of particles moving away from the shock without much scattering. 
The diffusion properties of particle ensembles are, however, statistical properties, i.e., the single trajectories may behave very differently from the average over many particles. The examples presented in Figure~\ref{fig:dw_traj} demonstrates that TAMSD can be very powerful diagnostic when dealing with poor particle statistics and/or heterogenous particle samples.  

\subsection{MSD and energy dependence}
\label{subsec:energy}

\begin{figure*}
    \includegraphics[width=\textwidth]{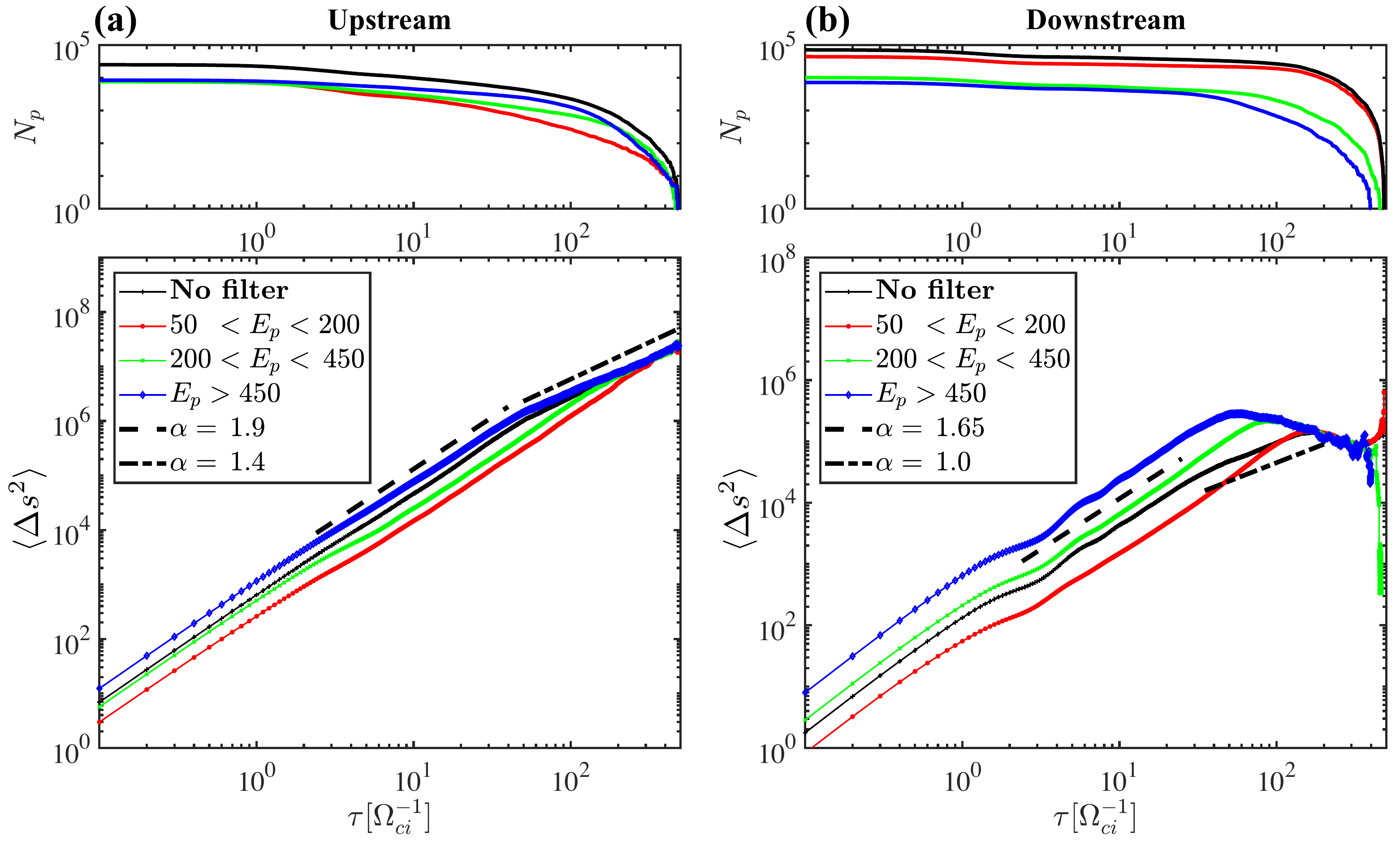}
    \caption{\textbf{(a)}, \emph{Bottom panel}: Upstream MSDs for increasing particles' energy (red, green and blue lines) and for the whole particle ensemble (black line). The dashed lines show the $\alpha$ exponents calculated in Section~\ref{subsec:upstream} and shown in Figure~\ref{fig:up_msd}. \textbf{(a)}, \emph{Top panel}: Number of particles in the upstream region as a function of time for different energies (same colors as bottom panel). \textbf{(b)}: same as \textbf{(a)}, but for the downstream region. The $\alpha$ exponents shown here (black dashed lines) are the ones obtained in Section~\ref{subsec:downstream} and presented in Figure~\ref{fig:dw_msd}. }
    \label{fig:fig_en}
\end{figure*}

{As discussed previously, the analyses presented so far treated upstream and downstream particles regardless of their energy, although the diffusion coefficient depends from the particles' energy (see Equation~\ref{eq:eq2}). Figure~\ref{fig:fig_en} shows upstream (a) and downstream (b) MSD calculations for particles in three different energy ranges. It should be noted that, since the shock transition zone is neglected in this study, and all the calculations are done in the local rest frame, the particles' energy remains roughly constant throughout the trajectory fragments analysed.}

{The dependence of the diffusion coefficient from the particles' energy in the upstream region is clearly found in Figure~\ref{fig:fig_en} (left panels), where the $\langle \Delta s^2 \rangle$ is found to be systematically larger for particles with larger energies. The black line, showing the upstream MSD for the whole particle ensemble, represents an `average' of what obtained in the different energy ranges.}

{Figure~\ref{fig:fig_en} shows that the MSD diagnostic for upstream particles is consistent with a superdiffusive scenario for all the energy ranges considered. It is worth noting that this result was predictable using single particle TAMSD upstream (demonstrating one of the strengths of TAMSD). In particular,  the histogram of single particle anomalous diffusion exponents reported in Figure~\ref{fig:up_hist_trj} shows that all the upstream particles have single particle diffusion exponent greater than one, indicating superdiffusion.}

{The transition from $\alpha = 1.9$ to $\alpha = 1.4$ is seen in all energy bands, although there is a trend to see this later at lower energies. The fact that the time at which the transition to $\alpha = 1.4$  happens changes is also influenced by a finite-box size selection effect. At later times, the results for the lowest energy particles becomes less significant because the number of particles in the ensemble decreases, as particles reencounter the shock.}

%\dt{}{It can also be noted that, at lower energies, the upstream particle MSD transitions later from the quasi-ballistic regime ($\alpha = 1.9$) to the $\alpha=1.4$ regime. We believe that this is a finite-box size selection effect, due to the fact that particles at lower energies are more likely to cross the shock transition towards the downstream. This effect has been tested with larger simulation domains, and does not influence our results, since we are not aiming for an exact calculation of the diffusion coefficient in this work.}

{In the shock downstream region, the dependence of the diffusion coefficient on the particle energy is also clear from Figure~\ref{fig:fig_en} (right hand panels). Once again, when the energy increases, the square displacement of particles becomes larger, and the MSD calculated for the entire particle ensemble represents an `average' between the behaviours observed at different energies.}

{When the energy filter is applied to the downstream particle ensemble, the regime corresponding to $\alpha \sim 1$  at later times (discussed in Section~\ref{subsec:downstream}), indicating normal diffusion, is not recovered. We suggest that this may be an effect related to the fact that the downstream particle ensemble is intrinsically heterogeneous.}

{Like for the upstream region, this analysis confirms what found using other diagnostics, namely that the shock downstream region has a much richer scenario as far as the particle transport properties are concerned.}

\section{Summary and conclusions}
\label{sec:conclusions}
Energetic particle transport regimes have been investigated in the framework of 1D, quasi-parallel shock hybrid PIC simulations. The simulation domain was divided into upstream and downstream regions, neglecting the transport regimes present within the shock transition. The transport regimes obtained for the two regions were found to have important differences. To the best of our knowledge, this is the first time that diffusion exponents have been extracted using particle trajectories advanced in the self-consistent fields produced with hybrid plasma simulations.

The usual technique to calculate anomalous diffusion exponents, based on the method of mean square displacement, relies on ensemble averages, and so it is very sensitive to poor particle statistics. It is often hard to track particles across the domain for long enough times in the framework of hybrid PIC simulations. Also, the dynamics of different particle populations are often very different in the shock simulation, and the ensemble average does not always capture this feature very well. For this reason, we have used a novel application of time-average mean square displacement together with the usual MSD method. TAMSD is a common diagnostic in scientific areas that have to deal with poor particle statistics, such as biophysics. It tries to overcome the problems of MSD by introducing a time averaging process on the single particle trajectories, and enables single particle estimates of the diffusion exponent.

In our simulation, we have found strong evidence for upstream superdiffusion. This evidence primarily arose from the calculation of MSD in the upstream frame for the reflected particles, in which we found extremely strong superdiffusion for intermediate times (with $\alpha \, \approx$ 1.9), and a superdiffusive behaviour with an $\alpha$ of 1.4 for long times. The EA TAMSD, which is a novel technique for plasma simulations, is in extremely good agreement with the more classic MSD, although there are some differences between the two diagnostics at intermediate times (6 to 45 $\Omega_{ci}^{-1}$), associated with transient behaviour. Single particle TAMSD enabled us to look at the statistical distribution of single particle anomalous diffusion exponents, in agreement with the other diagnostics. TAMSD is a powerful tool when we deal with poor particle statistics and with trajectory fragments of variable time extension.

Downstream of the shock, a much richer scenario of particle transport regimes is obtained. Both MSD and EA TAMSD show evidence for superdiffusion at intermediate times (6 to 30 $\Omega_{ci}^{-1}$), even though the anomalous diffusion exponents inferred using these two diagnostics are different (and equal to 1.65 and 1.85, respectively). A possible explanation for this difference in the diagnostics at intermediate times lies in the intrinsic heterogeneity of the particle sample. For later times, a diffusion exponent closer to unity, suggesting a tendency to Brownian motion is recovered using both MSD and EA TAMSD. When TAMSD is performed on single particle trajectories for the downstream sample, the single particle diffusion exponents are found to have a mean of 1.27 (consistent with moderate superdiffusion). However, the spread of the histogram of single particle $\alpha$ values is much larger than the one found for the upstream sample, ranging from highly subdiffusive to highly superdiffusive behaviours. We consider that this is the indication of different particle populations coexisting in the same region: from particles trapped in the shock fluctuations to particles moving away from the shock without interacting with the surrounding environment, (see examples presented in Figure~\ref{fig:dw_traj}).

We have presented evidence of anomalous diffusion of energetic ions in a single 1D simulation of a collisionless quasi-parallel shock. Such simulations obviously have limitations due to finite domain size and simulation time, so the applicability of our results to other shocks (in terms of shock and/or simulation parameters) remains an open question.

Future work will utilise fully 2D and 3D simulations \citep[][]{Trotta2019}, addressing diffusion in the magnetic field perpendicular and parallel directions. Moreover, it is possible to cross-correlate the diffusive regimes of different particles subsets with other important parameters, such as the energy gains, fluctuation levels, scattering times, or particle pitch angles and this will be the object of further investigation. We also intend to examine the transport regimes present in the shock transition layer, that have been neglected throughout this work, but which are probably crucial for understanding particle acceleration, especially at low energies. It is worth underlining that, throughout this work, we studied a shock with parameters compatible with those observed in the heliosphere. Further investigations will broaden the parameter space, in order to address whether these results are relevant for higher Mach number, astrophysical shocks. 

\section*{Acknowledgements}
DT acknowledges support of a studentship funded by the Perren Fund of the University of London. This research was supported by the UK Science and Technology Facilities Council (STFC) grant ST/P000622/1. This research utilised Queen Mary's Apocrita HPC facility, supported by QMUL Research-IT, http://doi.org/10.5281/zenodo.438045.

\bibliographystyle{mnras}
\interlinepenalty=10000
\bibliography{mybib} % if your bibtex file is called example.bib

\begin{thebibliography}{}
\makeatletter
\relax
\def\mn@urlcharsother{\let\do\@makeother \do\$\do\&\do\#\do\^\do\_\do\%\do\~}
\def\mn@doi{\begingroup\mn@urlcharsother \@ifnextchar [ {\mn@doi@}
  {\mn@doi@[]}}
\def\mn@doi@[#1]#2{\def\@tempa{#1}\ifx\@tempa\@empty \href
  {http://dx.doi.org/#2} {doi:#2}\else \href {http://dx.doi.org/#2} {#1}\fi
  \endgroup}
\def\mn@eprint#1#2{\mn@eprint@#1:#2::\@nil}
\def\mn@eprint@arXiv#1{\href {http://arxiv.org/abs/#1} {{\tt arXiv:#1}}}
\def\mn@eprint@dblp#1{\href {http://dblp.uni-trier.de/rec/bibtex/#1.xml}
  {dblp:#1}}
\def\mn@eprint@#1:#2:#3:#4\@nil{\def\@tempa {#1}\def\@tempb {#2}\def\@tempc
  {#3}\ifx \@tempc \@empty \let \@tempc \@tempb \let \@tempb \@tempa \fi \ifx
  \@tempb \@empty \def\@tempb {arXiv}\fi \@ifundefined
  {mn@eprint@\@tempb}{\@tempb:\@tempc}{\expandafter \expandafter \csname
  mn@eprint@\@tempb\endcsname \expandafter{\@tempc}}}

\bibitem[\protect\citeauthoryear{{Amano} \& {Hoshino}}{{Amano} \&
  {Hoshino}}{2007}]{Amano2007}
{Amano} T.,  {Hoshino} M.,  2007, \mn@doi [\apj] {10.1086/513599}, \href
  {http://adsabs.harvard.edu/abs/2007ApJ...661..190A} {661, 190}

\bibitem[\protect\citeauthoryear{{Amano} \& {Hoshino}}{{Amano} \&
  {Hoshino}}{2009}]{Amano2009}
{Amano} T.,  {Hoshino} M.,  2009, \mn@doi [\apj] {10.1088/0004-637X/690/1/244},
  \href {http://adsabs.harvard.edu/abs/2009ApJ...690..244A} {690, 244}

\bibitem[\protect\citeauthoryear{{Axford}, {Leer}  \& {Skadron}}{{Axford}
  et~al.}{1978}]{Axford1978}
{Axford} W.~I.,  {Leer} E.,   {Skadron} G.,  1978, in {Dergachev} V.~A.,
  {Kocharov} G.~E.,  eds, Cosmophysics. pp 125--134

\bibitem[\protect\citeauthoryear{Baring, Ellison  \& Jones}{Baring
  et~al.}{1995}]{Baring1995}
Baring M.,  Ellison D.,   Jones F.,  1995, \mn@doi [Advances in Space Research]
  {https://doi.org/10.1016/0273-1177(94)00124-J}, 15, 397

\bibitem[\protect\citeauthoryear{{Bell}}{{Bell}}{1978}]{Bell1978}
{Bell} A.~R.,  1978, \mn@doi [\mnras] {10.1093/mnras/182.2.147}, \href
  {http://adsabs.harvard.edu/abs/1978MNRAS.182..147B} {182, 147}

\bibitem[\protect\citeauthoryear{Blandford \& Eichler}{Blandford \&
  Eichler}{1987}]{Blandford1987}
Blandford R.,  Eichler D.,  1987, \mn@doi [Physics Reports]
  {https://doi.org/10.1016/0370-1573(87)90134-7}, 154, 1

\bibitem[\protect\citeauthoryear{{Blandford} \& {Ostriker}}{{Blandford} \&
  {Ostriker}}{1978}]{Blandford1978}
{Blandford} R.~D.,  {Ostriker} J.~P.,  1978, \mn@doi [\apjl] {10.1086/182658},
  \href {http://adsabs.harvard.edu/abs/1978ApJ...221L..29B} {221, L29}

\bibitem[\protect\citeauthoryear{Bovet, Fasoli, Ricci, Furno  \&
  Gustafson}{Bovet et~al.}{2015}]{Bovet2015}
Bovet A.,  Fasoli A.,  Ricci P.,  Furno I.,   Gustafson K.,  2015, \mn@doi
  [Phys. Rev. E] {10.1103/PhysRevE.91.041101}, 91, 041101

\bibitem[\protect\citeauthoryear{Bryant, Cline, Desai  \& McDonald}{Bryant
  et~al.}{1962}]{Bryant1962}
Bryant D.~A.,  Cline T.~L.,  Desai U.~D.,   McDonald F.~B.,  1962, \mn@doi
  [Journal of Geophysical Research] {10.1029/JZ067i013p04983}, 67, 4983

\bibitem[\protect\citeauthoryear{{Burgess} \& {Scholer}}{{Burgess} \&
  {Scholer}}{2015}]{burgess_book}
{Burgess} D.,  {Scholer} M.,  2015, {Collisionless Shocks in Space Plasmas}.
Cambridge University Press

\bibitem[\protect\citeauthoryear{Burnecki, Kepten, Garini, Sikora  \&
  Weron}{Burnecki et~al.}{2015}]{Burnecki2015}
Burnecki K.,  Kepten E.,  Garini Y.,  Sikora G.,   Weron A.,  2015, Scientific
  Reports, 5, 11306

\bibitem[\protect\citeauthoryear{Bykov, Ellison  \& Osipov}{Bykov
  et~al.}{2017}]{Bykov2017}
Bykov A.~M.,  Ellison D.~C.,   Osipov S.~M.,  2017, \mn@doi [Phys. Rev. E]
  {10.1103/PhysRevE.95.033207}, 95, 033207

\bibitem[\protect\citeauthoryear{{Caprioli}}{{Caprioli}}{2012}]{Caprioli2012}
{Caprioli} D.,  2012, \mn@doi [\jcap] {10.1088/1475-7516/2012/07/038}, \href
  {http://adsabs.harvard.edu/abs/2012JCAP...07..038C} {7, 038}

\bibitem[\protect\citeauthoryear{{Caprioli} \& {Spitkovsky}}{{Caprioli} \&
  {Spitkovsky}}{2014a}]{Caprioli2014a}
{Caprioli} D.,  {Spitkovsky} A.,  2014a, \mn@doi [\apj]
  {10.1088/0004-637X/783/2/91}, \href
  {http://adsabs.harvard.edu/abs/2014ApJ...783...91C} {783, 91}

\bibitem[\protect\citeauthoryear{Caprioli \& Spitkovsky}{Caprioli \&
  Spitkovsky}{2014b}]{caprioli2014c}
Caprioli D.,  Spitkovsky A.,  2014b, The Astrophysical Journal, 794, 47

\bibitem[\protect\citeauthoryear{Decker, Krimigis, Roelof, Hill, Armstrong,
  Gloeckler, Hamilton  \& Lanzerotti}{Decker et~al.}{2005}]{Decker2005}
Decker R.~B.,  Krimigis S.~M.,  Roelof E.~C.,  Hill M.~E.,  Armstrong T.~P.,
  Gloeckler G.,  Hamilton D.~C.,   Lanzerotti L.~J.,  2005, \mn@doi [Science]
  {10.1126/science.1117569}, 309, 2020

\bibitem[\protect\citeauthoryear{{Decker}, {Krimigis}, {Roelof}, {Hill},
  {Armstrong}, {Gloeckler}, {Hamilton}  \& {Lanzerotti}}{{Decker}
  et~al.}{2008}]{Decker2008}
{Decker} R.~B.,  {Krimigis} S.~M.,  {Roelof} E.~C.,  {Hill} M.~E.,  {Armstrong}
  T.~P.,  {Gloeckler} G.,  {Hamilton} D.~C.,   {Lanzerotti} L.~J.,  2008,
  \mn@doi [\nat] {10.1038/nature07030}, \href
  {http://adsabs.harvard.edu/abs/2008Natur.454...67D} {454, 67}

\bibitem[\protect\citeauthoryear{{Drury}}{{Drury}}{1983}]{Drury1983}
{Drury} L.~O.,  1983, \mn@doi [Reports on Progress in Physics]
  {10.1088/0034-4885/46/8/002}, \href
  {http://adsabs.harvard.edu/abs/1983RPPh...46..973D} {46, 973}

\bibitem[\protect\citeauthoryear{{Ellison}, {Moebius}  \&
  {Paschmann}}{{Ellison} et~al.}{1990}]{Ellison1990}
{Ellison} D.~C.,  {Moebius} E.,   {Paschmann} G.,  1990, \mn@doi [\apj]
  {10.1086/168544}, \href {http://adsabs.harvard.edu/abs/1990ApJ...352..376E}
  {352, 376}

\bibitem[\protect\citeauthoryear{Fisk}{Fisk}{2015}]{Fisk2015}
Fisk L.~A.,  2015, Journal of Physics: Conference Series, 642, 012009

\bibitem[\protect\citeauthoryear{{Gargat{\'e}} \& {Spitkovsky}}{{Gargat{\'e}}
  \& {Spitkovsky}}{2012}]{Gargate2012}
{Gargat{\'e}} L.,  {Spitkovsky} A.,  2012, \mn@doi [\apj]
  {10.1088/0004-637X/744/1/67}, \href
  {http://adsabs.harvard.edu/abs/2012ApJ...744...67G} {744, 67}

\bibitem[\protect\citeauthoryear{{Giacalone}}{{Giacalone}}{2004}]{Giacalone2004}
{Giacalone} J.,  2004, \mn@doi [\apj] {10.1086/421043}, \href
  {http://adsabs.harvard.edu/abs/2004ApJ...609..452G} {609, 452}

\bibitem[\protect\citeauthoryear{{Giacalone}}{{Giacalone}}{2012}]{Giacalone2012}
{Giacalone} J.,  2012, \mn@doi [\apj] {10.1088/0004-637X/761/1/28}, \href
  {http://adsabs.harvard.edu/abs/2012ApJ...761...28G} {761, 28}

\bibitem[\protect\citeauthoryear{{Giacalone} \& {Ellison}}{{Giacalone} \&
  {Ellison}}{2000}]{Giacalone2000}
{Giacalone} J.,  {Ellison} D.~C.,  2000, \mn@doi [\jgr] {10.1029/1999JA000018},
  \href {http://adsabs.harvard.edu/abs/2000JGR...10512541G} {105, 12541}

\bibitem[\protect\citeauthoryear{Giacalone, Burgess, Schwartz  \&
  Ellison}{Giacalone et~al.}{1992}]{Giacalone1992}
Giacalone J.,  Burgess D.,  Schwartz S.~J.,   Ellison D.~C.,  1992, \mn@doi
  [Geophysical Research Letters] {10.1029/92GL00379}, 19, 433

\bibitem[\protect\citeauthoryear{Golding \& Cox}{Golding \&
  Cox}{2004}]{Golding2004}
Golding I.,  Cox E.~C.,  2004, \mn@doi [Proceedings of the National Academy of
  Sciences] {10.1073/pnas.0404443101}, 101, 11310

\bibitem[\protect\citeauthoryear{H\"ofling \& Franosch}{H\"ofling \&
  Franosch}{2013}]{hafling2013}
H\"ofling F.,  Franosch T.,  2013, Reports on Progress in Physics, 76, 046602

\bibitem[\protect\citeauthoryear{{Jokipii}}{{Jokipii}}{1987}]{Jokipii1987}
{Jokipii} J.~R.,  1987, \mn@doi [\apj] {10.1086/165022}, \href
  {http://adsabs.harvard.edu/abs/1987ApJ...313..842J} {313, 842}

\bibitem[\protect\citeauthoryear{{Jones} \& {Ellison}}{{Jones} \&
  {Ellison}}{1991}]{Jones1991}
{Jones} F.~C.,  {Ellison} D.~C.,  1991, \mn@doi [\ssr] {10.1007/BF01206003},
  \href {http://adsabs.harvard.edu/abs/1991SSRv...58..259J} {58, 259}

\bibitem[\protect\citeauthoryear{{Kang} \& {Jones}}{{Kang} \&
  {Jones}}{1997}]{Kang1997}
{Kang} H.,  {Jones} T.~W.,  1997, \mn@doi [\apj] {10.1086/303646}, \href
  {http://adsabs.harvard.edu/abs/1997ApJ...476..875K} {476, 875}

\bibitem[\protect\citeauthoryear{{Kang}, {Ryu}  \& {Jones}}{{Kang}
  et~al.}{2017}]{Kang2017}
{Kang} H.,  {Ryu} D.,   {Jones} T.~W.,  2017, \mn@doi [\apj]
  {10.3847/1538-4357/aa6d0d}, \href
  {http://adsabs.harvard.edu/abs/2017ApJ...840...42K} {840, 42}

\bibitem[\protect\citeauthoryear{Kepten, Bronshtein  \& Garini}{Kepten
  et~al.}{2013}]{Kepten2013}
Kepten E.,  Bronshtein I.,   Garini Y.,  2013, \mn@doi [Phys. Rev. E]
  {10.1103/PhysRevE.87.052713}, 87, 052713

\bibitem[\protect\citeauthoryear{Kepten, Weron, Sikora, Burnecki  \&
  Garini}{Kepten et~al.}{2015}]{Kepten2015}
Kepten E.,  Weron A.,  Sikora G.,  Burnecki K.,   Garini Y.,  2015, \mn@doi
  [PLOS ONE] {10.1371/journal.pone.0117722}, 10, 1

\bibitem[\protect\citeauthoryear{Klafter, Blumen  \& Shlesinger}{Klafter
  et~al.}{1987}]{Klafter1987}
Klafter J.,  Blumen A.,   Shlesinger M.~F.,  1987, \mn@doi [Phys. Rev. A]
  {10.1103/PhysRevA.35.3081}, 35, 3081

\bibitem[\protect\citeauthoryear{{Krymskii}}{{Krymskii}}{1977}]{Krymskii1977}
{Krymskii} G.~F.,  1977, Akademiia Nauk SSSR Doklady, \href
  {http://adsabs.harvard.edu/abs/1977DoSSR.234.1306K} {234, 1306}

\bibitem[\protect\citeauthoryear{{Kucharek}, {Scholder}  \&
  {Matthews}}{{Kucharek} et~al.}{2000}]{Kucharek2000}
{Kucharek} H.,  {Scholder} M.,   {Matthews} A.~P.,  2000, Nonlinear Processes
  in Geophysics, \href {http://adsabs.harvard.edu/abs/2000NPGeo...7..167K} {7,
  167}

\bibitem[\protect\citeauthoryear{{Malkov}}{{Malkov}}{1997}]{Malkov1997}
{Malkov} M.~A.,  1997, \mn@doi [\apj] {10.1086/304990}, \href
  {http://adsabs.harvard.edu/abs/1997ApJ...491..584M} {491, 584}

\bibitem[\protect\citeauthoryear{Matthews}{Matthews}{1994}]{Matthews1994}
Matthews A.~P.,  1994, \mn@doi [Journal of Computational Physics]
  {https://doi.org/10.1006/jcph.1994.1084}, 112, 102

\bibitem[\protect\citeauthoryear{Meerschaert \& Scalas}{Meerschaert \&
  Scalas}{2006}]{Meerschaert2006}
Meerschaert M.~M.,  Scalas E.,  2006, \mn@doi [Physica A: Statistical Mechanics
  and its Applications] {https://doi.org/10.1016/j.physa.2006.04.034}, 370, 114

\bibitem[\protect\citeauthoryear{Metzler \& Klafter}{Metzler \&
  Klafter}{2000}]{Metzler2000}
Metzler R.,  Klafter J.,  2000, \mn@doi [Physics Reports]
  {https://doi.org/10.1016/S0370-1573(00)00070-3}, 339, 1

\bibitem[\protect\citeauthoryear{Metzler, Jeon, Cherstvy  \& Barkai}{Metzler
  et~al.}{2014}]{metzler2014}
Metzler R.,  Jeon J.-H.,  Cherstvy A.~G.,   Barkai E.,  2014, \mn@doi [Phys.
  Chem. Chem. Phys.] {10.1039/C4CP03465A}, 16, 24128

\bibitem[\protect\citeauthoryear{{\noopsort{Nes}{van Nes}}, Reinhard,
  Sanderson, Wenzel  \& Zwickl}{{\noopsort{Nes}{van Nes}}
  et~al.}{1984}]{nes1984}
{\noopsort{Nes}{van Nes}} P.,  Reinhard R.,  Sanderson T.~R.,  Wenzel K.-P.,
  Zwickl R.~D.,  1984, \mn@doi [Journal of Geophysical Research: Space Physics]
  {10.1029/JA089iA04p02122}, 89, 2122

\bibitem[\protect\citeauthoryear{{Perri} \& {Zimbardo}}{{Perri} \&
  {Zimbardo}}{2007}]{Perri2007}
{Perri} S.,  {Zimbardo} G.,  2007, \mn@doi [\apjl] {10.1086/525523}, \href
  {http://adsabs.harvard.edu/abs/2007ApJ...671L.177P} {671, L177}

\bibitem[\protect\citeauthoryear{{Perri} \& {Zimbardo}}{{Perri} \&
  {Zimbardo}}{2008}]{Perri2008a}
{Perri} S.,  {Zimbardo} G.,  2008, \mn@doi [Journal of Geophysical Research
  (Space Physics)] {10.1029/2007JA012695}, \href
  {http://adsabs.harvard.edu/abs/2008JGRA..113.3107P} {113, A03107}

\bibitem[\protect\citeauthoryear{{Perri} \& {Zimbardo}}{{Perri} \&
  {Zimbardo}}{2009a}]{Perri2009a}
{Perri} S.,  {Zimbardo} G.,  2009a, \mn@doi [Advances in Space Research]
  {10.1016/j.asr.2009.04.017}, \href
  {http://adsabs.harvard.edu/abs/2009AdSpR..44..465P} {44, 465}

\bibitem[\protect\citeauthoryear{{Perri} \& {Zimbardo}}{{Perri} \&
  {Zimbardo}}{2009b}]{Perri2009b}
{Perri} S.,  {Zimbardo} G.,  2009b, \mn@doi [\apjl]
  {10.1088/0004-637X/693/2/L118}, \href
  {http://adsabs.harvard.edu/abs/2009ApJ...693L.118P} {693, L118}

\bibitem[\protect\citeauthoryear{{Perri} \& {Zimbardo}}{{Perri} \&
  {Zimbardo}}{2012a}]{Perri2012}
{Perri} S.,  {Zimbardo} G.,  2012a, \mn@doi [\apj]
  {10.1088/0004-637X/750/2/87}, \href
  {http://adsabs.harvard.edu/abs/2012ApJ...750...87P} {750, 87}

\bibitem[\protect\citeauthoryear{{Perri} \& {Zimbardo}}{{Perri} \&
  {Zimbardo}}{2012b}]{Perri2012_2}
{Perri} S.,  {Zimbardo} G.,  2012b, \mn@doi [\apj] {10.1088/0004-637X/754/1/8},
  \href {http://adsabs.harvard.edu/abs/2012ApJ...754....8P} {754, 8}

\bibitem[\protect\citeauthoryear{{Perri} \& {Zimbardo}}{{Perri} \&
  {Zimbardo}}{2015}]{Perri2015}
{Perri} S.,  {Zimbardo} G.,  2015, \mn@doi [\apj] {10.1088/0004-637X/815/1/75},
  \href {http://adsabs.harvard.edu/abs/2015ApJ...815...75P} {815, 75}

\bibitem[\protect\citeauthoryear{{Perri}, {Zimbardo}, {Effenberger}  \&
  {Fichtner}}{{Perri} et~al.}{2015}]{Perri2015b}
{Perri} S.,  {Zimbardo} G.,  {Effenberger} F.,   {Fichtner} H.,  2015, \mn@doi
  [\aap] {10.1051/0004-6361/201425295}, \href
  {http://adsabs.harvard.edu/abs/2015A%26A...578A...2P} {578, A2}

\bibitem[\protect\citeauthoryear{Prete, Perri  \& Zimbardo}{Prete
  et~al.}{2019}]{Prete2019}
Prete G.,  Perri S.,   Zimbardo G.,  2019, \mn@doi [Advances in Space Research]
  {https://doi.org/10.1016/j.asr.2019.01.002}

\bibitem[\protect\citeauthoryear{Qian, Sheetz  \& Elson}{Qian
  et~al.}{1991}]{quian1991}
Qian H.,  Sheetz M.,   Elson E.,  1991, \mn@doi [Biophysical Journal]
  {https://doi.org/10.1016/S0006-3495(91)82125-7}, 60, 910

\bibitem[\protect\citeauthoryear{Ramos-Fern{\'a}ndez, Mateos, Miramontes,
  Cocho, Larralde  \& Ayala-Orozco}{Ramos-Fern{\'a}ndez
  et~al.}{2004}]{Fernandez2004}
Ramos-Fern{\'a}ndez G.,  Mateos J.~L.,  Miramontes O.,  Cocho G.,  Larralde H.,
    Ayala-Orozco B.,  2004, \mn@doi [Behavioral Ecology and Sociobiology]
  {10.1007/s00265-003-0700-6}, 55, 223

\bibitem[\protect\citeauthoryear{Reynolds}{Reynolds}{2008}]{reynolds2008}
Reynolds S.~P.,  2008, \mn@doi [Annual Review of Astronomy and Astrophysics]
  {10.1146/annurev.astro.46.060407.145237}, 46, 89

\bibitem[\protect\citeauthoryear{{Riquelme} \& {Spitkovsky}}{{Riquelme} \&
  {Spitkovsky}}{2011}]{Riquelme2011}
{Riquelme} M.~A.,  {Spitkovsky} A.,  2011, \mn@doi [\apj]
  {10.1088/0004-637X/733/1/63}, \href
  {http://adsabs.harvard.edu/abs/2011ApJ...733...63R} {733, 63}

\bibitem[\protect\citeauthoryear{Saxton}{Saxton}{2012}]{Saxton2012}
Saxton M.~J.,  2012, \mn@doi [Biophysical Journal]
  {https://doi.org/10.1016/j.bpj.2012.10.038}, 103, 2411

\bibitem[\protect\citeauthoryear{Scholer \& Terasawa}{Scholer \&
  Terasawa}{1990}]{Scholer1990}
Scholer M.,  Terasawa T.,  1990, \mn@doi [Geophysical Research Letters]
  {10.1029/GL017i002p00119}, 17, 119

\bibitem[\protect\citeauthoryear{Scholer, Ipavich, Gloeckler  \&
  Hovestadt}{Scholer et~al.}{1982}]{Scholer1982}
Scholer M.,  Ipavich F.~M.,  Gloeckler G.,   Hovestadt D.,  1982, \mn@doi
  [Journal of Geophysical Research: Space Physics] {10.1029/JA088iA03p01977},
  88, 1977

\bibitem[\protect\citeauthoryear{Scholer, Kucharek  \& Giacalone}{Scholer
  et~al.}{2000}]{Scholer2000}
Scholer M.,  Kucharek H.,   Giacalone J.,  2000, \mn@doi [Journal of
  Geophysical Research: Space Physics] {10.1029/1999JA000324}, 105, 18285

\bibitem[\protect\citeauthoryear{{Servidio}, {Haynes}, {Matthaeus}, {Burgess},
  {Carbone}  \& {Veltri}}{{Servidio} et~al.}{2016}]{Servidio2016}
{Servidio} S.,  {Haynes} C.~T.,  {Matthaeus} W.~H.,  {Burgess} D.,  {Carbone}
  V.,   {Veltri} P.,  2016, \mn@doi [Physical Review Letters]
  {10.1103/PhysRevLett.117.095101}, \href
  {http://adsabs.harvard.edu/abs/2016PhRvL.117i5101S} {117, 095101}

\bibitem[\protect\citeauthoryear{{Strauss} \& {Effenberger}}{{Strauss} \&
  {Effenberger}}{2017}]{Strauss2017}
{Strauss} R.~D.~T.,  {Effenberger} F.,  2017, \mn@doi [\ssr]
  {10.1007/s11214-017-0351-y}, \href
  {http://adsabs.harvard.edu/abs/2017SSRv..212..151S} {212, 151}

\bibitem[\protect\citeauthoryear{Sugiyama}{Sugiyama}{2011}]{Sugiyama2011a}
Sugiyama T.,  2011, \mn@doi [Physics of Plasmas] {10.1063/1.3552026}, 18,
  022302

\bibitem[\protect\citeauthoryear{{Sugiyama} \& {Shiota}}{{Sugiyama} \&
  {Shiota}}{2011}]{Sugiyama2011}
{Sugiyama} T.,  {Shiota} D.,  2011, \mn@doi [\apjl]
  {10.1088/2041-8205/731/2/L34}, \href
  {http://adsabs.harvard.edu/abs/2011ApJ...731L..34S} {731, L34}

\bibitem[\protect\citeauthoryear{Sundberg, Haynes, Burgess  \&
  Mazelle}{Sundberg et~al.}{2016}]{Sundberg2016}
Sundberg T.,  Haynes C.~T.,  Burgess D.,   Mazelle C.~X.,  2016, The
  Astrophysical Journal, 820, 21

\bibitem[\protect\citeauthoryear{{Trotta} \& {Burgess}}{{Trotta} \&
  {Burgess}}{2019}]{Trotta2019}
{Trotta} D.,  {Burgess} D.,  2019, \mn@doi [\mnras] {10.1093/mnras/sty2756},
  \href {http://adsabs.harvard.edu/abs/2019MNRAS.482.1154T} {482, 1154}

\bibitem[\protect\citeauthoryear{{Trotta} \& {Zimbardo}}{{Trotta} \&
  {Zimbardo}}{2015}]{Trotta2015}
{Trotta} E.~M.,  {Zimbardo} G.,  2015, \mn@doi [Journal of Plasma Physics]
  {10.1017/S0022377814000592}, \href
  {http://adsabs.harvard.edu/abs/2015JPlPh..81a3208T} {81, 325810108}

\bibitem[\protect\citeauthoryear{{\noopsort{Weeren}{van Weeren}},
  {R{\"o}ttgering}, {Br{\"u}ggen}  \& {Hoeft}}{{\noopsort{Weeren}{van Weeren}}
  et~al.}{2010}]{vanweeren2010}
{\noopsort{Weeren}{van Weeren}} R.~J.,  {R{\"o}ttgering} H.~J.~A.,
  {Br{\"u}ggen} M.,   {Hoeft} M.,  2010, \mn@doi [Science]
  {10.1126/science.1194293}, \href
  {http://adsabs.harvard.edu/abs/2010Sci...330..347V} {330, 347}

\bibitem[\protect\citeauthoryear{{\noopsort{Weeren}{van Weeren}}
  et~al.,}{{\noopsort{Weeren}{van Weeren}} et~al.}{2017}]{vanweeren2017}
{\noopsort{Weeren}{van Weeren}} R.~J.,  et~al., 2017, \mn@doi [Nature
  Astronomy] {10.1038/s41550-016-0005}, \href
  {http://adsabs.harvard.edu/abs/2017NatAs...1E...5V} {1, 0005}

\bibitem[\protect\citeauthoryear{Weron, Burnecki, Akin, Sol{\'{e}}, Balcerek,
  Tamkun  \& Krapf}{Weron et~al.}{2017}]{Weron2017}
Weron A.,  Burnecki K.,  Akin E.~J.,  Sol{\'{e}} L.,  Balcerek M.,  Tamkun
  M.~M.,   Krapf D.,  2017, \mn@doi [Scientific Reports]
  {10.1038/s41598-017-05911-y}, 7, 5404

\bibitem[\protect\citeauthoryear{{Wolff} \& {Tautz}}{{Wolff} \&
  {Tautz}}{2015}]{Wolff2015}
{Wolff} M.,  {Tautz} R.~C.,  2015, \mn@doi [\aap]
  {10.1051/0004-6361/201525907}, \href
  {http://adsabs.harvard.edu/abs/2015A%26A...580A..58W} {580, A58}

\bibitem[\protect\citeauthoryear{Wong, Gardel, Reichman, Weeks, Valentine,
  Bausch  \& Weitz}{Wong et~al.}{2004}]{Wong2004}
Wong I.~Y.,  Gardel M.~L.,  Reichman D.~R.,  Weeks E.~R.,  Valentine M.~T.,
  Bausch A.~R.,   Weitz D.~A.,  2004, \mn@doi [Phys. Rev. Lett.]
  {10.1103/PhysRevLett.92.178101}, 92, 178101

\bibitem[\protect\citeauthoryear{{Zimbardo} \& {Perri}}{{Zimbardo} \&
  {Perri}}{2013}]{Zimbardo2013}
{Zimbardo} G.,  {Perri} S.,  2013, \mn@doi [\apj] {10.1088/0004-637X/778/1/35},
  \href {http://adsabs.harvard.edu/abs/2013ApJ...778...35Z} {778, 35}

\bibitem[\protect\citeauthoryear{{Zimbardo} \& {Perri}}{{Zimbardo} \&
  {Perri}}{2017}]{Zimbardo2017}
{Zimbardo} G.,  {Perri} S.,  2017, \mn@doi [Nature Astronomy]
  {10.1038/s41550-017-0163}, \href
  {http://adsabs.harvard.edu/abs/2017NatAs...1E.163Z} {1, 0163}

\bibitem[\protect\citeauthoryear{{Zimbardo} \& {Perri}}{{Zimbardo} \&
  {Perri}}{2018}]{Zimbardo2018}
{Zimbardo} G.,  {Perri} S.,  2018, \mn@doi [\mnras] {10.1093/mnras/sty1438},
  \href {http://adsabs.harvard.edu/abs/2018MNRAS.478.4922Z} {478, 4922}

\makeatother
\end{thebibliography}

\newpage
\appendix

\section{Testing TAMSD with Monte Carlo models}
\label{appendix}

Although TAMSD is a common diagnostic used to address anomalous transport properties of many biological systems, it has never been used in the framework of numerical simulations of collisionless plasmas. On the other hand, self-consistent trajectories from hybrid PIC shock simulations have rarely been used to address the diffusion regimes at play in the plasma environment. Therefore, we have tested TAMSD on particle trajectories obtained using a simpler, Monte Carlo model for anomalous transport. In the Monte Carlo model, particles are advanced integrating a Langevin type equation in the form of a stochastic differential equation \citep[e.g.][]{Strauss2017}:
\begin{equation}
\label{eq:a1}
    dx_i = v_\mathrm{ran} dt_i .
\end{equation}
The distribution of $v_\mathrm{ran}$ controls the diffusion regime, that is, scattering episodes for the particles are modelled through the distribution of $v_\mathrm{ran}$. The random velocity is changed randomly at each scattering time $t_i$. In the case of superdiffusion, the scattering times $t_i$ have a power-law-tailed probability distribution. The Monte Carlo model employed to generate the trajectories used for testing the anomalous diffusion diagnostics has extensively been used to bridge observations and models of anomalous transport in plasmas both for collisionless shocks and plasma turbulence \citep[][]{Prete2019}.

In this model, it is possible to set the exponent $\mu$ of the power-law distribution of scattering times. The $\mu$ exponent is related to the anomalous diffusion exponent through the relation \citep[e.g.][]{Klafter1987}: 
\begin{equation}
    \alpha =  4- \mu .
\end{equation}
The particle scattering times are then defined through the distribution 
\begin{equation}
\label{eq:A3}
\psi(t) = 
	\begin{cases}
	A   & t \leq t_0\\
	A \left(\frac{t}{t_0} \right)^{-\mu} & t > t_0
	\end{cases} \: ,
\end{equation}
where $A$ is a normalisation constant and $t_0$ is a free parameter. The details about this model can be found in \citet[][]{Zimbardo2013, Trotta2015}.

It is possible to compare the theoretical values for $\alpha$ with those obtained from the particle trajectories analysis using MSD, TAMSD and EA TAMSD. It is worth noting that for the self-consistent plasma model, the theoretical $\alpha$ is inferred from the analysis of particle trajectories (i.e., it is not a free parameter).

The first test uses a diffusion exponent $\alpha = 1$, corresponding to normal (Brownian) diffusion. The equations of motion (Equation \ref{eq:a1}) are integrated for an ensemble of 10$^4$ particles, so that the ensemble averages do not suffer from statistical noise. The final time for the particle trajectories is of 36000 time units with a timestep of unit time.
\begin{figure*}
    \includegraphics[width=\textwidth]{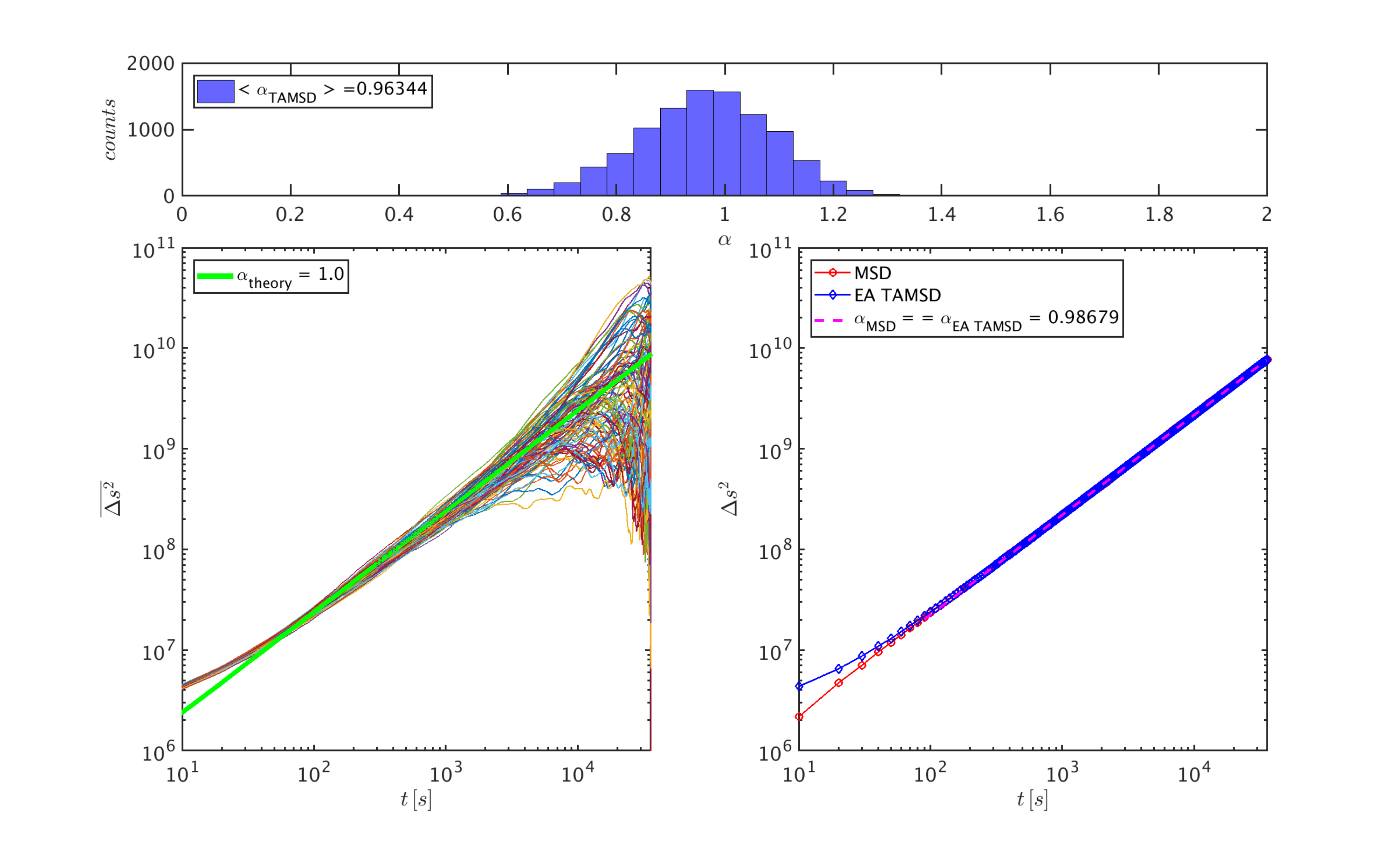}
    \caption{Top: Single particle diffusion exponents histogram, calculated using TAMSD. Bottom Left: A sample of single particle TAMSD curves and their comparison with the theoretical diffusion exponent $\alpha_\mathrm{theory} =1$. Bottom right: MSD (red line) and EA TAMSD (blue line) curves for the particle sample. The dashed magenta line represents the $\alpha$ exponent obtained fitting MSD and EA TAMSD.}
    \label{fig:a1}
\end{figure*}
Results are presented in Figure ~\ref{fig:a1}. Results for MSD, TAMSD and EA TAMSD are consistent and capture the expected theoretical $\alpha$ used in the simulation. The single particle TAMSD exponents have been calculated by fitting the single particle TAMSD curves (Figure ~\ref{fig:a1}, bottom-left) in the range of times before they start to become noisy. The slight underestimation of $\alpha$ using TAMSD ($\alpha_\mathrm{TAMSD} \sim 0.96$ vs $\alpha_\mathrm{theory} = 1.0$) is an effect from which the TAMSD diagnostic is known to suffer. We note that MSD and EA TAMSD capture the value of the expected $\alpha$ with a smaller error than single particle TAMSD and, moreover, they return the same value when $\alpha$ is fitted ($\alpha_\mathrm{MSD} = \alpha_\mathrm{EA \, TAMSD} \sim 0.99$). The extremely strong agreement between MSD and EA TAMSD also has a theoretical explanation: when normal diffusion is achieved, the central limit theorem is satisfied, so time and ensemble averages must converge for asymptotic times. 

\begin{figure*}
    \includegraphics[width=\textwidth]{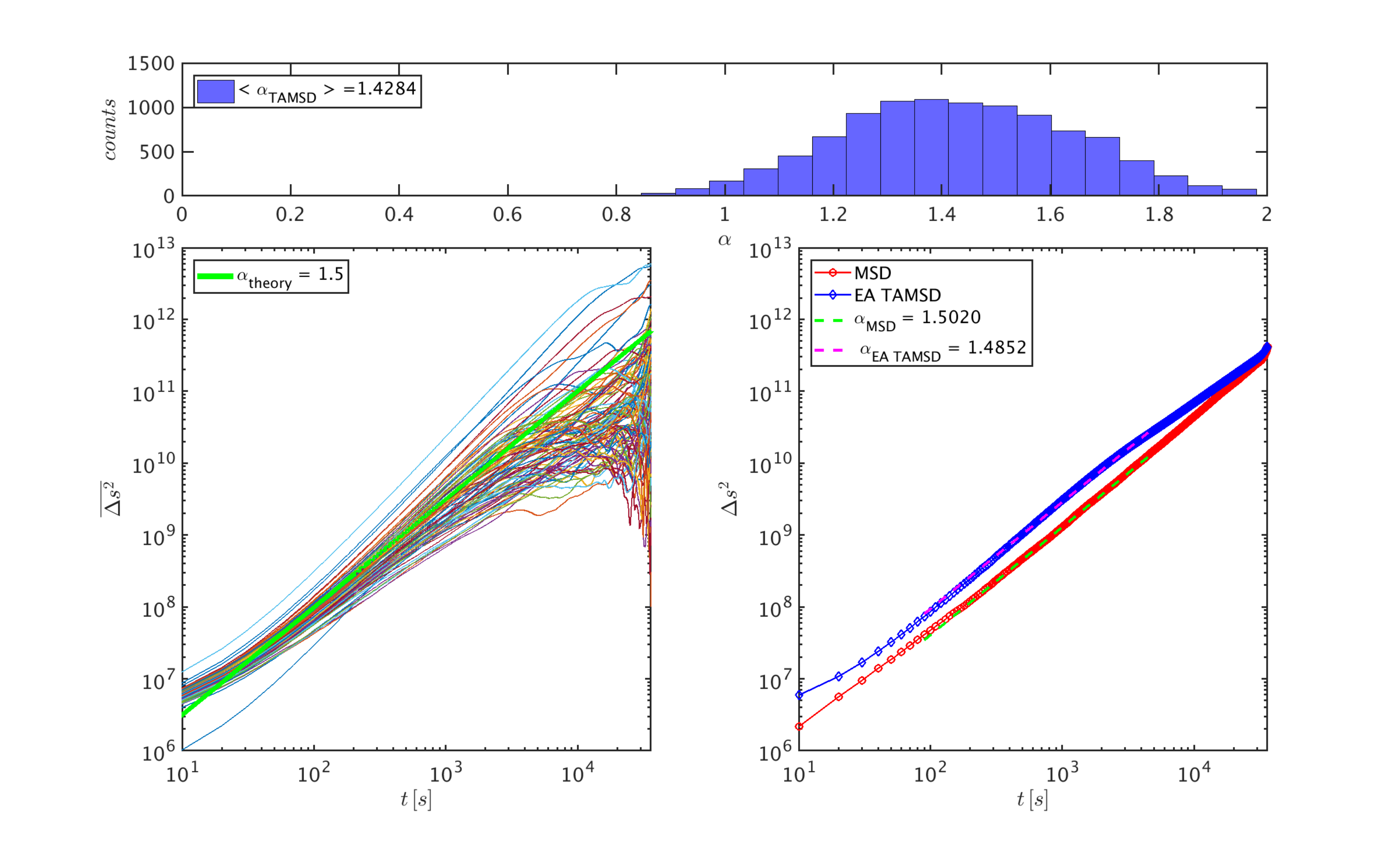}
    \caption{Top: Single particle diffusion exponents histogram, calculated using TAMSD. Bottom Left: A sample of single particle TAMSD curves and their comparison with the theoretical diffusion exponent $\alpha_\mathrm{ theory} =1.5$. Bottom right: MSD (red line) and EA TAMSD (blue line) curves for the particle sample. The dashed lines are obtained by fitting the MSD (green) and the EA TAMSD (magenta) in the time range indicated.}
    \label{fig:a2}
\end{figure*}

For comparison, a superdiffusive case has been analysed as well by means of the Monte Carlo method. Here, an anomalous diffusion exponent $\alpha$ of 1.5 (i.e., $\mu$=2.5) has been imposed in Equation \ref{eq:A3}. The number of trajectories followed and the final simulation time are the same as the previous case. Figure ~\ref{fig:a2} presents the results of the test. The single particle diffusion exponents have been fitted from single particle TAMSDs with the criterion described above. The mean single particle diffusion exponent $\langle\alpha_\mathrm{TAMSD}\rangle$ value is 1.43 and slightly underestimates the theoretical value of 1.5. Again, this is a known effect for the TAMSD diagnostic. However, the inferred $\langle\alpha_\mathrm{TAMSD}\rangle$ is in good agreement with the theoretical value, as can be also seen in the histogram shown in the top panel of Figure~\ref{fig:a2}. 

The diffusion exponents obtained using MSD and EA TAMSD are both consistent with the expected $\alpha_\mathrm{theory}$ within reasonable errors, although their slopes have been fitted for a shorter time window. The slight discrepancy between the MSD and the EA TAMSD curves, as well as the $\sim$ 1\% difference in the calculated $\alpha$ values may be due to the fact that in the superdiffusive regime the central limit theorem does not hold anymore, and so we would expect some difference when taking time and ensemble averages.

The results of the tests are summarised in Table \ref{tab:a1}. It is worth noting that the diagnostics used to infer the anomalous diffusion exponents are consistent, but show slight differences. However, TAMSD is a flexible and powerful tool to use when dealing with datasets affected by poor particle statistics and trajectories with varying duration. The tests presented in this appendix help us understand the robustness of the diagnostics presented.
\begin{table*}
 \caption{Summary of anomalous diffusion exponents calculated with MSD, TAMSD and EA TAMSD for Monte Carlo simulations with assigned $\alpha$}
 \label{tab:a1}
 \begin{tabular}{llll}
  \hline
  $\alpha_\mathrm{theory}$ & $\alpha_\mathrm{MSD}$ & $\alpha_\mathrm{EA \, TAMSD}$ & $\langle\alpha_\mathrm{TAMSD}\rangle$ \\
  \hline
  1.0 & 0.99 & 0.99 & 0.96\\[2pt] % additional height spacing for enhanced symbol legibility
  1.50 & 1.50 & 1.48 & 1.43\\[2pt]
  \hline
 \end{tabular}
\end{table*}

%%%%%%%%%%%%%%%%%%%%%%%%%%%%%%%%%%%%%%%%%%%%%%%%%%



%\bibliographystyle{mnras}

% Don't change these lines
\bsp	% typesetting comment
\label{lastpage}
\end{document}